  \documentclass{aa}
\usepackage{txfonts,graphicx,lscape}
\begin{document}
\title {Carbon abundances of early B-type stars in the solar vicinity\thanks{Based on 
 observations collected at the European Southern Obser\-vatory, Chile, ESO 074.B-0455(A).}}
\subtitle{Non-LTE line-formation for \ion{C}{ii/iii/iv} and self-consistent atmospheric parameters}   
\author {M. F. Nieva\inst{1,2} \and N. Przybilla\inst{1}}
\offprints{nieva@sternwarte.uni-erlangen.de}
\institute{Dr. Remeis Sternwarte Bamberg. Sternwartstr. 7, D-96049 Bamberg,
Germany
\and Observat\'orio Nacional, Rua General Jos\'e Cristino 77 CEP 20921-400, Rio de Janeiro, Brazil}
\date{Received... ; accepted ... }
\abstract
{Precise determinations of the chemical composition in early B-type stars constitute fundamental
observational constraints on 
stellar and
galactochemical evolution. Carbon, in particular, is one of the most abundant metals in 
the Universe but analyses in early-type stars are known to show
inconclusive results. Large discrepancies between analyses of different
lines in \ion{C}{ii}, a failure to establish the \ion{C}{ii/iii} ionization balance and 
the derivation of systematically lower abundances than from other indicators
like \ion{H}{ii} regions and young FG-type stars pose long-standing
problems.}
{We discuss improvements to the non-LTE modelling of the visual 
line spectrum and to the spectral analysis of early B-type stars, as well as their
consequences for stellar parameter and abundance derivations. The most relevant
sources of systematic uncertainies and their effects on the analysis are investigated.
Consequences for the present-day carbon abundance in the solar vicinity are discussed.}
{We present a comprehensive and robust \ion{C}{ii/iii/iv} model for non-LTE line-formation
calculations based on carefully selected atomic data. The model is calibrated
with high-S/N spectra of six apparently slow-rotating early B-type dwarfs and giants, which 
cover a wide parameter range and are randomly distributed in the solar neighbourhood. 
A self-consistent quantitative spectrum analysis is performed using an extensive
iteration scheme to determine stellar atmospheric parameters
and to select the appropriate atomic data used for the derivation of chemical
abundances.}
{We establish the carbon ionization balance for all sample stars based on a
unique set of input atomic data. Consistency is achieved for all modelled carbon lines 
of the sample stars.
Highly accurate atmospheric parameters and a homogeneous carbon abundance of 
$\log$\,(C/H)\,$+$\,12\,$=$\,8.32\,$\pm$\,0.04 with reduced systematic errors are derived. 
Present evolution models for massive stars indicate that this value may require only
a small adjustment because of the effects of rotational mixing, 
by $<$\,$+$0.05\,dex per sample star. 
This results in a present-day stellar carbon abundance
in the solar neighbourhood, which is in good agreement with recent
determinations of the solar value and with the gas-phase abundance of the
Orion \ion{H}{ii} region. Our finding of a
homogeneous present-day carbon abundance also conforms with predictions of
chemical-evolution models for the Galaxy.
Moreover, the present approach allows us to constrain the effects of systematic errors 
on fundamental parameters and abundances. This suggests that most of the difficulties 
found in previous work may be related to large systematic effects in the atmospheric
parameter determination and/or inaccuracies in the atomic data.
}
{}
\keywords{Line: formation -- Stars: %atmospheres, 
early type, fundamental parameters, abundances -- Galaxy: abundances, solar neighbourhood}
\titlerunning {Non-LTE line formation for \ion{C}{ii-iv} in early B-type stars}
\authorrunning {Nieva \& Przybilla}
\maketitle
\section {Introduction} 
Studies of the chemical composition of early-type stars
provide valuable observational constraints to our understanding of broad
fields like stellar evolution, nucleosynthesis and galactochemical evolution.
The accuracy of elemental abundance determinations defines the degree to
which theory can be tested. However, quantitative spectroscopy relies on
many model assumptions itself. %Stellar atmospheres need to be modelled: 
A detailed understanding of the interaction between the radiation
field and the plasma in the stellar atmosphere 
is thus required. Any weakness in the physical
model limits the analysis even when high
quality in observation and data reduction~is~achieved.

Concerning Galactochemical evolution (e.g. models of Hou et~al.~\cite{Houetal00}; 
Chiappini et al.~\cite{Chiappinietal01},~\cite{Chiappinietal03}), 
early-type stars~are important probes for the current chemical
composition of~the Galaxy. They provide spatial information of  
elemental abundances also beyond the solar vicinity, allowing Galactic metallicity gradients 
to be derived 
(e.g. Gummersbach et al.~\cite{Gummersbachetal98}; Rolleston et al.~\cite{Rollestonetal00};
Daflon \& Cunha~\cite{DaCu04}). Abundances from young stars complement data
from \ion{H}{ii} regions, allowing the current status of
the nucleosynthesis of metals in the course of the cosmic cycle of matter to be constrained.
Pristine abundances can be derived not only for Galactic early-type stars, 
but also for objects in the metal-poor environments of the Magellanic Clouds,
using the present generation of large telescopes and high-resolution spectrographs (e.g. Korn 
et al.~\cite{Kornetal02}, \cite{Kornetal05}; Rolleston et al.~\cite{Rollestonetal03}; Hunter et
al.~\cite{Hunteretal07}). Such investigations can be extended to more
distant galaxies of the Local Group, and even beyond, by analysing
intermediate-resolution spectra of early-type supergiants 
(e.g. Trundle et al.~\cite{Trundleetal02}; Urbaneja et al.~2005ab).
Moreover, CNO abundances derived from quantitative spectral analyses of
massive early-type stars provide fundamental observational constraints 
for stellar evolution models. Basic stellar parameters and 
abundance patterns of the light elements resulting from mixing with nuclear-processed 
matter facilitates an empirical evaluation of %the quality of 
different evolution models (e.g.
Heger \& Langer~\cite{HeLa00}; Maeder \& Meynet~\cite{MaMe00}).

Carbon plays a special r\^ole among the light elements. It is a
primary element 
created in the fundamental 3$\alpha$-process and as such it provides the seed
for the subsequent synthesis of all heavier elements
(Burbidge et al.~\cite{B2FH}; Cameron~\cite{Cameron57}). Carbon is an essential
catalyst for the nucleosynthesis of H into He through the 
CNO cycle in massive and intermediate-mass stars. 
The element also constitutes the basis of all organic chemistry. 

Carbon abundances derived from early-type stars were the subject 
of numerous spectroscopic studies in the past decades, 
from the pioneering work of Uns\"old~(\cite{unsold42}) to numerous others since then.
The quality of observed data and the complexity of the model
calculations and spectral analysis have been improved with great effort over these years, 
however some inconsistencies remained until the present.
The 'classical' carbon problem in early-type stars addresses the extremely low 
abundances derived from the strong \ion{C}{ii} $\lambda$4267\,{\AA} doublet when 
compared with those derived from weaker lines, 
as pointed out by Kane et al.~(\cite{kane80}), referring to work by 
Hardorp \& Scholz~(\cite{hardorp}) and Kodaira \& Scholz~(\cite{kodaira}). 
Later on, more complications were found, such as discrepancies from both 
strong \ion{C}{ii} $\lambda\lambda$6578/82\,{\AA} and 4267\,{\AA} multiplets,  
$\lambda\lambda$6578/82\,{\AA} and the weak lines, among the \ion{C}{ii} weak lines themselves
and in abundances derived from \ion{C}{ii} and \ion{C}{iii}.

In nearby, bright and apparently slow-rotating stars it is possible to
 derive carbon abundances from weak lines which are less sensitive to the details 
 of the analysis. However, studies of extragalactic
or fast-rotating stars are restricted to the stronger lines
because of S/N-constraints or rotational smearing. In those cases, one
 might be interested
to analyse not only the \ion{C}{ii} $\lambda\lambda$6578/82\,{\AA} and 
the 4267\,{\AA} multiplets, but also other strong 
\ion{C}{iii} or \ion{C}{iv} lines, in order to assure ionization equilibrium.

A crucial step in the spectral modelling was to abandon
the approximation of local thermodynamic equilibrium (LTE) in line-formation
calculations and allow for deviations (non-LTE). Several model atoms 
have been discussed in literature
(Lennon~\cite{lennon83}; Sakhibullin~\cite{sak87}; Eber \& Butler~\cite{ebbut88};
Grigsby et al.~\cite{Grigsbyetal92};
Sigut~\cite{Sigut96}; Przybilla et al.~\cite{n01}; Lanz \& Hu\-beny~\cite{lh03}, \cite{lh07}). 
The model atom of Eber \& Butler found wide application for 
abundance analyses of mostly unevolved early-type stars in the solar
neighbourhood 
(e.g. Gies \& Lambert~\cite{gl92}; Kilian~\cite{k92}; Cunha \&
Lambert~\cite{cl94}; Gummersbach et al.~\cite{Gummersbachetal98}; 
Andrievsky et al.~\cite{Andrievskyetal99}; 
Daflon et al.~\cite{daf99},~2001ab). 

The \ion{C}{ii}\,$\lambda$4267\,{\AA} multiplet poses a challenge to non-LTE line-formation
 calculations (e.g. Lambert~\cite{Lambert93}; Sigut~\cite{Sigut96}). 
 Even in non-LTE studies this and the $\lambda\lambda$6578/82\,{\AA} multiplet
 failed to reproduce observation consistently (Grigsby et
al.~\cite{Grigsbyetal92}; Hunter et al.~\cite{Hunteretal07}).
Systematically lower abundances are derived than from 
other weaker \ion{C}{ii} lines (Gies \& Lambert~\cite{gl92}), as in the LTE case.
Observed
trends for \ion{C}{iii} lines may also be poorly matched by non-LTE model
calculations (Grigsby et al.~\cite{Grigsbyetal92}). A failure to establish the  \ion{C}{ii/iii} ionization
equilibrium in non-LTE was also reported. Differences in abundance from the two ions can amount up to a factor $\sim$5--10 (Daflon et al.~\cite{daf01b}; Hunter et al.~\cite{Hunteretal07}).
There are even more inconsistencies with published carbon abundances
from early-type stars in a broader context that require other explanation,
as most of the published studies avoid the lines sensitive to non-LTE effects.

A comparison of the available studies of early B-type stars indicates
that present-day carbon abundances in the solar vicinity are highly inhomogeneous 
(even for stars within a single cluster) and largely sub-solar. 
This is in contrast to the findings of a uniform abundance in the gas-phase 
of the interstellar medium (ISM) within 1.5\,kpc of the Sun 
(Sofia \& Meyer~\cite{SoMe01}, and references therein).
It cannot be understood from current stellar and Galactochemical evolution
models either.
Young massive stars that form out of a molecular cloud within a short timescale 
can be expected to show a homogeneous carbon 
abundance\footnote{Carbon experiences a slight depletion in
the course of the CNO cycle, which may become observable 
at the stellar surface because of rotational mixing, see 
Sect.~\ref{neighbourhood} for a discussion.},
which also coincides with that of their surrounding \ion{H}{ii} nebula. Moreover, this
abundance is expected to be higher than that of objects from previous
generations of star formation in their neighbourhood, like young ($\leq$2\,Gyr) 
F and G stars, and that of the Sun. In reality, significant systematic 
differences exist (see e.g. Sofia \& Meyer~\cite{SoMe01};
Herrero~\cite{Herrero03}), shedding doubt on the reliability of carbon
abundances derived from early B-type stars.

Quantitative spectroscopic studies depend on the quality of the
observational data, the physical approximations made in the modelling 
approach (stellar atmospheres,
model atoms) and the analysis technique. The analysis procedure also
involves several steps (data reduction, stellar parameter determination
and abundance derivation), which are all critical for the quality of the final results.
It is difficult to control systematic effects in the whole process.
Therefore, the sources of discrepant results from the literature are hard to be identified.

One underestimated problem in the derivation of the chemical composition is 
certainly the selection of the input atomic data for non-LTE line-formation calculations
(see e.g. Si\-gut~\cite{Sigut96} for \ion{C}{ii}). Even when the model atmosphere
computations reproduce the plasma conditions in the stellar
photosphere realistically, use of different atomic data for the
line-formation calculations can result in large systematic
errors in the final abundances. 
The structure of atoms and therefore the atomic line spectra are governed by the 
laws of quantum mechanics, i.e. they are essentially independent of the 
environment the atoms are located in.
This property can be used to calibrate model atoms for non-LTE calculations, 
by demanding that the model should reproduce observation reliably 
over the entire parameter space of relevance. In practice, this means that model
atoms need to be tested for a few cases of well-established plasma conditions.
The atmospheres of early B-type stars far from the Eddington limit provide 
such testbeds, as they can be described well by rather simple physical models.
A small sample of stars, spanning the relevant range in spectral type and
luminosity, is sufficient for such tests.

The present study aims to solve several inconsistencies concerning the 
derivation of carbon abundances in early B-type stars and to apply the new modelling
and analysis technique in stars of the solar vicinity.
It focuses on both i) the critical selection of appropriate atomic input data 
 from {\em ab-initio} calculations and approximation formulae and
 ii) a self-consistent derivation of the atmospheric parameters of
the sample stars from ionization equilibria. This detailed and reliable study 
allows us to derive abundances with unprecedented accuracy for a 
so far small but very well constrained sample of stars covering strategic 
points of the stellar parameter space and  randomly 
distributed in OB associations and in the field of the solar neighbourhood.

This study is part of our efforts to improve on the accuracy of
quantitative spectroscopy in early B-type stars
(Nieva \& Przybilla~\cite{np07a}: Paper~1; Nieva \&
Przybilla~\cite{np06}: Paper~2) 
and related objects (Przybilla et al.~\cite{Przybillaetal05}, 2006a-c).
The hybrid non-LTE approach (LTE model atmospheres\,$+$\,non-LTE line formation) 
was shown to be adequate for analyses of OB-type main
sequence and giant stars in Paper~1, allowing highly consistent results to be derived.
In particular, practically all measurable H and He lines from the Balmer 
jump to the near-IR (where available) have been reproduced with a single model for each of the present 
sample stars, meeting a crucial boundary condition for all further analyses.
Paper~2 discusses a solution to the 'classical' carbon problem, i.e. discrepant carbon abundances from 
 \ion{C}{ii} $\lambda$4267\,{\AA} and weaker lines, and  also a solution to the notorious discrepancy of the prominent $\lambda\lambda$6578/82\,{\AA} and the 4267\,{\AA} multiplets.
There, consistency was achieved for up to 21~\ion{C}{ii} features of our sample stars. 
Those results are part of the present work, which discusses the topic in a much broader~context.
Here, we present for the first time a quantitative agreement of abundances from \ion{C}{ii} and \ion{C}{iii} lines for this kind of star. We also study \ion{C}{iv} lines and establish the \ion{C}{ii/iii/iv} ionization equilibrium in a self-consistent way, never considered before 
in similar work. The linelist rises up to 40 features. In the present work we investigate  
the systematic errors arising due to uncertainties in the atomic data and stellar parameters 
and quantify the difference between LTE and non-LTE analyses, for the first time in this extent.
Implications on effective temperature scales for abundance determitations are discussed. This extensive study provides us with reliable carbon abundances for a discussion in the context of galactochemical evolution.

The paper is organised as follows: Section~2 describes the model calculations including 
a description of the carbon model. Section~3 describes the empirical 
model calibration via an extensive and self-consistent iteration and the 
sensitivity of carbon line-formation calculations to atomic data and 
atmospheric parameter variations. Section~4 summarises the final results for parameters
and carbon abundances of individual lines for the sample stars. In Sect.~5 we provide
a comparison of our results to previous studies and the conclusions on the
present-day carbon abundance in the solar neighbourhood are discussed in
Sect.~6. A summary is given in Sect.~7.\\[-4mm]

\section {Model calculations}

The non-LTE line-formation computations for carbon follow the
methodology discussed in our previous study of the hydrogen and helium
spectra in early-type stars (Paper 1).
A hybrid non-LTE approach is employed to solve
the restricted non-LTE problem on the basis of prescribed LTE atmospheres.
This technique provides an efficient way to compute realistic synthetic
spectra in all cases where the atmospheric structure is close to LTE. 
The computational efforts can thus be concentrated on robust
non-LTE line-formation calculations.\\[-5.5mm]

\subsection {Model atmospheres and programs}

The model atmospheres are computed with the {\sc Atlas9}~code (Kurucz~\cite{kur93b}),
 which assumes plane-parallel geometry,~chemical homogeneity and
hydrostatic, radiative and local thermo\-dynamic equilibrium (LTE). Line
blanketing is realised via consideration of Opacity Distribution Functions (ODFs,
Ku\-rucz~\cite{kur93a}).
We adopt standard solar abundances (Grevesse \& Sauval \cite{gs98}) in all 
computations, as in Paper 1. Helium abundance and
microturbulence are adjusted according to the results from the quantitative
analysis of the programme stars -- requiring some iteration, as explained
below.
The model atmospheres are held fixed in the non-LTE calculations.

Non-LTE level populations and model spectra are obtained with
recent versions of {\sc Detail} and {\sc Surface} (Gid\-dings~\cite{gid81}; Butler \&
Giddings~\cite{but_gid85}; both updated by K. Butler). The coupled~radiative 
transfer and statistical
equilibrium equations are~solved with {\sc Detail}, employing an Accelerated 
Lambda Iteration scheme of Rybicki \&
Hummer~(\cite{rh91}). This allows even complex ions to be treated in a
realistic way. 
Continuous opacities due to hydrogen and helium (for actual abundances) are considered in
non-LTE, line blocking is accounted for in LTE via Kurucz' ODFs.
Synthetic spectra are calculated with {\sc Surface}, using refined line-broadening theories.
Microturbulence is consistently accounted for in both steps with {\sc
Detail} and {\sc Surface}.
 
Non-LTE level populations for hydrogen and \ion{He}{i/ii} are computed using 
recent model atoms by Przybilla \& Butler~(\cite{pb04}) and Przybilla~(\cite{p05}),
 respectively.
The synthetic H and \ion{He}{i/ii} lines give excellent agreement with
observations, in the optical as well as in the near-IR (Paper 1).
This is a prerequisite for modelling metal lines which overlap with the
(broad) hydrogen or helium features. In particular, the 
\ion{C}{ii}\,$\lambda\lambda$6578/82\,{\AA} doublet is affected in the
present case. \\[-5.5mm]

\subsection {The \ion{C}{ii-iv} model atom}\label{atom}

 We give a short summary of the input atomic data for the construction of the
\ion{C}{ii/iii/iv} model atom in this section.
A more detailed description of our motivation for choosing these atomic
data will be given in Section 3.2, as this turned out to be
critical for the realistic modelling of the observed spectra.

{\bf \ion{C}{ii}.~} This model ion 
considers $LS$-coupled terms up to principal quantum number
$n$\,$=$\,10 and angular momentum $\ell$\,$=$\,9 (66 levels)
explicitly in the non-LTE calculations, with all fine-structure sub-levels 
combined into one. Additional levels up to $n$\,$=$\,14 are computed in LTE 
relative to the ground state of \ion{C}{iii}. Level energies are
adopted from Moore~(\cite{Moore93}), Sigut~(\cite{Sigut96}) and
Quinet~(\cite{Quinet98}). The doublet and quartet spin
systems are treated simultaneously.
Oscillator strengths ($gf$-values) from three sources are considered:
fine-structure data from {\em ab-initio} computations using the
multiconfiguration
Hartree-Fock method in the Breit-Pauli approximation of Froese Fischer \&
Tachiev~(\cite{FFT04}, FFT04), data from application of the Breit-Pauli
{\bf {\emph R}}-matrix method (Na\-har~\cite{nahar02a}, N02a) and results
obtained in the
Opacity Project (OP) from the {\bf {\emph R}}-matrix method 
assuming LS-coupling 
(Yan, Taylor \& Seaton~\cite{yan87}). Our primary source of $gf$-values is FFT04,
followed by OP and N02a for the remaining
transitions. Inter\-combination transitions are neglected because of very
small oscillator strengths. 
Photoionizations cross-sections are adopted from
the OP (Yan \& Seaton~\cite{yansea87}) for levels up to $n$\,$=$\,9
and $\ell$\,$=$\,3 with a correction of the threshold frequencies to observed
values and from Nahar \& Pradhan~(\cite{nahar97}:~NP97) for the remainder.
Effective collision strengths for electron impact excitation among
the lowest 16 LS-states are adopted from {\bf {\emph R}}-matrix computations of Wilson,
Bell \& Hudson~(\cite{WBH05}, \cite{WBH07}). An empirical increase by a
factor two was applied to the 3s\,$^2$S--3p\,$^2$P$^{\rm o}$ data (see
Sect.~\ref{atomic_sensitivity}).
Collisional excitation for transitions without detailed data are treated using
the Van Regemorter~(\cite{van62}) approximation -- in the optically
allowed case -- and via the semi-empirical Allen~(\cite{allen73}) formula in
the optically forbidden case. Collision strengths $\Omega$ varying between 0.01
($\Delta n\,\geq\,4$) to 100 ($\Delta n\,=\,0$) are employed, as suggested by
evaluation of the detailed data from {\em ab-initio} computations of Wilson
et al.~(\cite{WBH05}, \cite{WBH07}).
Collisional ionization rates are evaluated according to the
Seaton~(\cite{seaton62})
approximation. Threshold photoionization cross-sections are adopted 
from OP and NP97, allowing for an empirical correction of one order of
magnitude higher for the 6f\,$^2$F$^{\circ}$
and 6g\,$^2$G levels -- corresponding to the upper levels of the
C\,{\sc ii} $\lambda\lambda$6151 and 6462\,{\AA} transitions, respectively.

{\bf \ion{C}{iii}.~} This model accounts for $LS$-terms up to 
$n$\,$=$\,7 and  $\ell$\,$=$\,7 (70 levels)
explicitly in the statistical equilibrium calculations. In a similar way to \ion{C}{ii},
 levels up to $n$\,$=$\,14 are computed 
in LTE relative to the ground state of the next ionization stage. The
two spin systems (singlet and triplet) are treated simultaneously.
Level energies are taken from the NIST
database\footnote {http://physics.nist.gov/PhysRefData/ASD/indices.html}
and energies for 9 of the highest levels are adopted from N02a. 
Two sources are considered for oscillator strengths:
N02a and additional values from Eber~(\cite{eb87}).
Intercombination transitions are also implemented when their values are 
non-negligible ($f$\,$>$\,$10^{-4}$). 
Photoionization cross-sections are taken from S. Nahar's 
webpage\footnote{http://www-astronomy.mps.ohio-state.edu/$\sim$nahar/px.html}.
Maxwellian-averaged collision strengths for electron impact excitation
among the lowest 24 terms are adopted from the
{\bf {\emph R}}-matrix computations of Mitnik et al.~(\cite{mit03}).
Collisional excitation for the remaining transitions and collisional
ionization are treated in analogy to the \ion{C}{ii} ion, using appropriate $gf$-values
and threshold photoionization cross-sections.

\begin{figure*}
\centering
\includegraphics[width=0.74\linewidth]{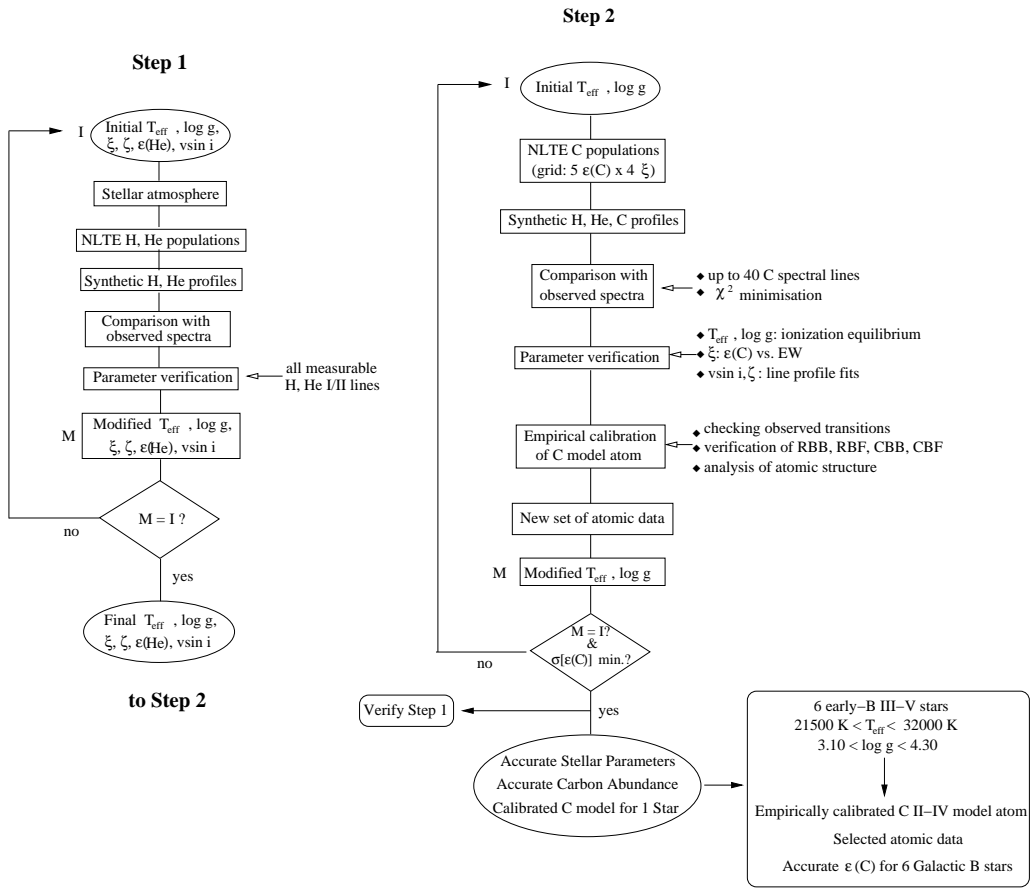}
\caption{Flux diagram of our extensive iterative procedure for the carbon model atom calibration and the simultaneous precise determination of stellar parameters, as applied to each programme star.
{\sc rbb/cbb}: radiative/collisional bound-bound, {\sc rbf/cbf}:
radiative/collisional bound-free transition data, {\sc ew}: equivalent width.
Step~1 is discussed in Paper~1, the sensitivity of C lines to different sets of atomic data in Sect.~\ref{atomic_sensitivity} and the sensitivity of C to parameter variations in Sect.~\ref{param_variations} of the present paper. See the text for~details.}
\label{hhec}
\end{figure*}

{\bf \ion{C}{iv}.~} $LS$-terms up to 
$n$\,$=$\,10 and $\ell$\,$=$\,9 (53 levels)
are treated explicitly. Additional levels up to $n$\,$=$\,14 are 
computed in LTE relative to the ground state of \ion{C}{v}. 
Oscillator strengths from {\it ab-initio} calculations 
using the Breit-Pauli 
{\bf {\emph R}}-matrix method (Nahar~\cite{nahar02b}) are adopted. 
Photoionization cross-sections are also taken from Nahar's webpage.
Effective collision strengths for electron impact excitation of transitions among 
the lowest 24 fine-structure levels are taken from Aggarwal \& Keenan
(\cite{agg04}) and subsequently co-added.
All remaining transitions, as well as collisional ionization, 
are treated in analogy to \ion{C}{ii}.\\[-3mm]

The resulting \ion{C}{ii/iii/iv} model atom accounts for more than 1300
radiative and more than 5300 collisional transitions, $\sim$200 $LS$-coupled energy levels 
and over 20\,000 frequency points -- the latter allow the detailed resonance 
structure of the photoionization cross-sections to be sampled with $<$$0.5$$~\AA$~resolution.
Accuracies of the atomic data can range from a typical 10-20\% for
{\em ab-initio} computations to orders of magnitude for approximation formulae.
Finally, Voigt profiles are adopted in the formal solution using 
{\sc Surface}. Wavelengths and oscillator strengths of most of the observed transitions are 
taken from Wiese et al.~(\cite{wiese96}). 
For \ion{C}{ii} $\lambda\lambda$6151.3/5 and 6461.9\,{\AA} the 
 wavelengths are adopted from Kurucz \& Bell (\cite{KuBe95}) and the $f$-values from N02a. 
Radiative damping parameters are calculated from OP lifetimes and coefficients 
for collisional broadening by electron impact are adopted from
Griem~(\cite{griem74}, for the C\,{\sc ii} $\lambda$4267\,{\AA} doublet) or
computed according to Cowley (\cite{cowley71}). Detailed tabulations from 
quantummechanical computations for Stark broadening of several \ion{C}{iv} 
transitions (Sch\"oning~\cite{sch93}) are also used.
The spectral lines used for abundance analysis are listed
later in Table~3.\\[-5.5mm]

\section {Simultaneous \ion{C}{ii/iii/iv} model-atom calibration and atmospheric parameter determination}\label{calib}

 A reliable set of atomic data considered in model atoms for non-LTE 
calculations can be selected by demanding that the model reproduce the observations over the
 entire stellar parameter space of relevance. This procedure can be regarded as a 'calibration'
 and requires 
precise atmospheric parameters for the test sample stars. 
These should be free of systematic errors 
in order to prevent one from being mislead when optimising the selection of input atomic
data. Unfortunately, the parameters of stars are {\em a priori} not known, they
also need to be inferred from observations. 
Consequently, a {\em simultaneous} solution for an optimal set of input atomic data
and all stellar parameters is required. These parameters should also allow the hydrogen and helium spectra to be reproduced.

A sample of six apparently slow-rotating B-type dwarfs and giants (spectral class B0-B2, 
 luminosity class V-III) randomly distributed in
 Galactic OB associations and in the field of the solar vicinity are taken as calibration stars. 
The observational data consists of 
high-resolution spectra with broad wavelength coverage at very high-S/N (up to
 $\sim$800 in the Johnson $B$-band),
obtained with FEROS on the ESO 2.2m telescope (La Silla, Chile), see
Paper~1 for~details.

\subsection {Extensive iteration on stellar and atomic variables}

The atmospheric parameter derivation and selection of input atomic data
are simultaneously performed in an extensive iteration process. 
When possible, we try to separate the effects of the fundamental parameters 
and the atomic data on the synthetic spectra -- the basis for the comparison
with observation -- in order to achieve a better understanding of the problem. 
This is facilitated by boundary conditions, like the ionization balance
(all ionization stages of an element are required to indicate the same abundance) 
or the rules and regularities of atomic physics.

The iteration is performed on effective temperature $T_\mathrm{eff}$ and surface 
gravity $\log g$, as well as micro-, macroturbulent and projected rotational 
velocities ($\xi$, $\zeta$ and $v\sin i$, respectively), helium
and carbon abundances (hereafter $\varepsilon(\mathrm{He})$\footnote{using
the standard logarithmic scale
$\varepsilon(\mathrm{X})=\log(\mathrm{X/H})+12$} and 
$\varepsilon(\mathrm{C})$,
respectively) and different sets of atomic data. Only the metallicity is fixed
to a standard solar value (Grevesse \& Sauval~\cite{gs98}), a not too
critical assumption which is furthermore validated {\em a posteriori}
(Przybilla et al.~2007, in prep.). This comprises an enormous 
number of variables (atmospheric parameters and atomic data) in the iterative scheme
summarised as a flux diagram in Fig.~\ref{hhec}. 
The first step concerning the H/He spectrum is solved in Paper~1. There, the
\ion{He}{i/ii} ionization equilibrium is the main indicator for $T_\mathrm{eff}$ (for 
the hotter stars), \emph{all} Balmer lines for 
$\log g$, the \ion{He}{ii} lines for $\xi$ and all He lines for $\zeta$ and $v\sin i$.
The second step, involving carbon, is required for a fine tuning of the 
atmospheric parameter
determination since the metal lines are more sensitive to parameter variations 
than the H and He lines. Therefore it is possible to derive them with a better 
precision than only from hydrogen and helium but at the same time consistently within the 
error limits. Effective temperature and $\log g$ are refined
by establishing the \ion{C}{ii/iii/iv} ionization equilibrium in the
hottest stars and the \ion{C}{ii/iii} ionization balance in the cooler stars.
The microturbulent velocity is inferred in the standard way by 
demanding the carbon
abundances of the individual lines to be independent of equivalent width. 
Macroturbulent and projected
rotational velocities are determined by detailed fitting of the carbon line
profiles. Line fits are performed on the basis of small grids of synthetic
spectra with different $\xi$ and $\varepsilon(\mathrm{C})$ via $\chi^2$-minimisation. 

\begin{figure}
\centering
% \sidecaption
\includegraphics[width=.99\linewidth]{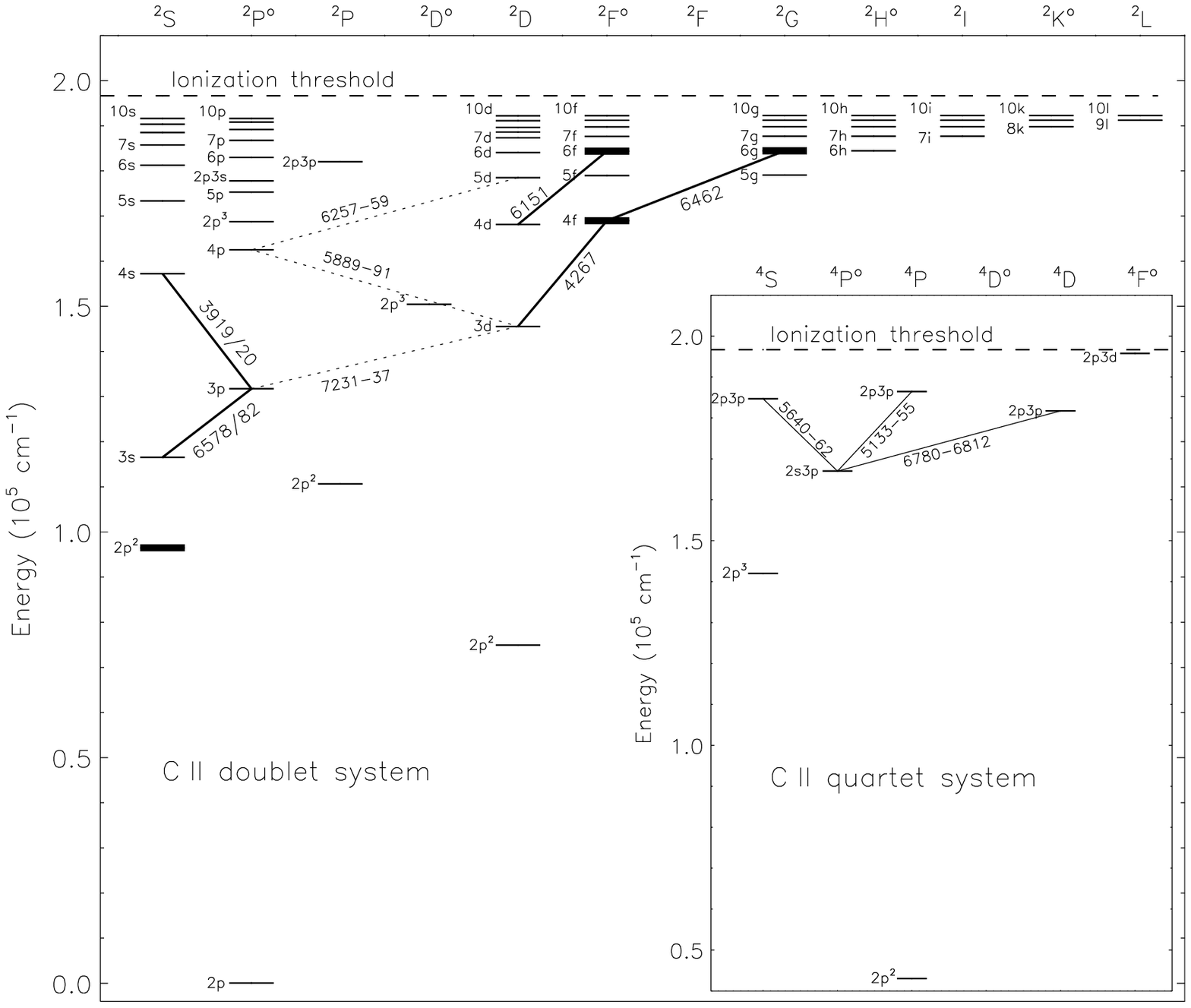}
\caption{Grotrian diagram for the \ion{C}{ii} doublet and quartet spin systems. 
 The observed multiplets in our spectra analysed here are identified. 
\emph {Levels}: those marked in bold correspond to levels discussed in 
Figs.~\ref{RBFlevel}~and~\ref{CBF}.
 \emph {Multiplet transitions}: those marked by thick lines are highly sensitive 
to variations of input data for photoionization and collisional cross-sections and
for collisional excitation
(discussed in Figs.~\ref{RBFlines}, \ref{CBF} and \ref{CBB}, respectively).
The latter and those marked by thin lines are considered in our linelist 
(Table~3) for abundance derivation. 
Those marked by dotted lines are excluded from the analysis 
because of contamination with telluric lines ($\lambda\lambda$6257-59,
7231-37\,{\AA})
or they are too weak even at low $T_\mathrm{eff}$ ($\lambda$5889-91\,{\AA}).
Nevertheless they are accounted for in the calculations of the level populations.
}
\label{grotrian}
\end{figure}

Since different sets of atomic data are available we have, first of all, to intercompare them
and judge their reliability to mini\-mise the uncertainties in the $\varepsilon$(C)-determination. 
Starting from an initial model atom we use our expertise for stepwise improvement, guided by a careful 
analysis of the atomic structure of the model ions and the reactions of the 
spectrum synthesis to parameter variations. Over 100 models were built until 
consistency with all sample stars was achieved simultaneously.
 Some examples of the effects from different model atom realisations on the spectrum synthesis are given in the next section.
 
The atmospheric parameters derived from C ionization equilibrium in Step 
2 are verified by re-iterating Step 1 as a final check for consistency.
Excellent agree\-ment with the observed H and He spectra 
is obtained in all cases, simultaneously in the visual and the near-IR
(where available). In addition, consistency of the observed and model 
spectral energy distributions (Paper 2) is found 
for the final set of stellar parameters.
  
By application of the procedure to all programme stars it is possible to calibrate the 
\ion{C}{ii-iv} model over the entire parameter range 
(21\,500 $\le$ $T_\mathrm{eff}$ $\le$ 32\,000 K, 3.10 $\le$ $\log g$ $\le$
4.30, for dwarfs and giants), resulting in a final reference set of atomic data. 
Note that the high-quality spectra available to us are essential for this
success. They allow us to analyse a wide variety of C lines,
many of which have never been considered before in the study of early B
stars. Some weak lines turned out to be highly sensitive
to non-LTE effects and/or to atmospheric parameter variations, 
namely the \ion{C}{iv} lines and \ion{C}{ii}\,$\lambda\lambda$6151 and
6462\,{\AA}, which change from absorption to emission at higher temperatures 
(discussed in Paper 2). The reproduction of the observed trends, despite
this high sensitivity, puts strong constraints on the robustness of the final
model atom. Note also that the presence of lines from three C
ionization stages, in hotter stars of our sample, allows $T_\mathrm{eff}$
and $\log g$ to be derived from the ionization equilibrium alone, independent
of other indicators. 
The strength of this calibration lies in the 
simultaneous analysis of 
the large number of C lines of different ionization stages in stars 
covering a wide parameter range. In this way we are able to constrain a final set of 
atomic data {\em independent} of any specific stellar atmosphere environment.
This reference model can be used in further applications with a 
simplified iterative scheme, where the only remaining variables are the atmospheric
parameters and the C~abundance.\\[-5.5mm]

\subsection {Sensitivity of carbon lines to atomic data}\label{atomic_sensitivity}

Reliable level populations are a prerequisite for an accurate non-LTE analysis.
They can be obtained only when: 
{\sc i})~the local temperatures and particle densities are known
(i.e. the atmospheric structure); 
{\sc ii}) the radiation field is realistic; 
{\sc iii}) all relevant processes in the statistical equilibrium equations are taken into account
{\em and} 
{\sc iv}) high-quality atomic data are available.
In particular, {\sc i}) and {\sc ii}) require a realistic physical model of the
stellar atmosphere (see Paper~1 for a discussion of the hybrid non-LTE
approach) and an accurate atmospheric parameter
determination. Points {\sc iii}) and {\sc iv}) are related to the model
atoms for the non-LTE calculations.
Shortcomings in any of {\sc i})--{\sc iv}) result in
increased uncertainties/errors of the analysis. 

Due to the interdependency of all transitions (over 6000 in this case) and 
the non-local character of the radiation field, even a restricted non-LTE problem like the
one investigated here is highly complex. It is
impossible to quantify {\em a priori} the sensitivity of the spectral lines
to variations of some of the atomic input data. 
Therefore, one of the few remaining reasons for the large spread of carbon
abundances found in literature (see Sect.~\ref{comparison}) may be different
realisations of model atoms (levels/transitions considered, atomic
data, approximations). Choosing an optimum set of input
atomic data is not trivial and the construction of reliable model
atoms for non-LTE calculations requires a calibration, 
guided by extensive comparisons
with observation. In the following we summarise the experiences made
in this process.

The comparison of observation with model spectra for \ion{C}{iii} and 
\ion{C}{iv}, as computed with our initial model ions, reveals little need for improvement
in the studied stellar parameter range.
Both ions are relatively simple, showing (earth)alkali electron
configurations, which pose little challenge to {\em ab-initio} computations.
However, the actual choice of radiative and collisional data turned out
to be a critical factor for line-formation computations of non-LTE-sensitive
transitions in \ion{C}{ii}, which is known to be problematic from the
literature (e.g. Lambert~\cite{Lambert93}; Sigut~\cite{Sigut96}). 
This was shortly summarised in Paper~2 but not discussed in detail.
Here we provide a few compa\-risons of atomic data available from
the literature. The influence of atomic data on selected \ion{C}{ii} lines
is also addressed.
These transitions are highlighted in the Grotrian diagram of our \ion{C}{ii} model 
(Fig.~\ref{grotrian}), which will help us to illustrate some of the channels
leading to the marked non-LTE sensitivity. 
\\ [-2.5mm]

\emph{Photoionization cross-sections.} 
The strength of a spectral line 
can be strongly influenced by photoionizations which may impact
 level populations decisively.
On the other hand, photoionization rates $R_{ij}$ depend implicitly on the level
populations.
The largest contribution to the integral $R_{ij}$ comes from frequencies 
with large flux and large cross-section. The flux maximum in early B stars is located 
longward of the Lyman jump.
The ionization of \ion{C}{ii} is essentially determined by the
rates from the highly-populated ground state and the low-excitation levels. 
The relative importance of the low-excitation levels may be strengthened in cases
where the ground state ionization potential coincides with that of a major
opacity contributor, \ion{He}{i} in the present case. Photoionizations from the ground state may then
be less efficient because of the reduced stellar flux shortward of the
ionization edge. Recombinations, on the other hand,
are important for the population of high-excitation levels (preferentially
at high $\ell$, i.e. states with large statistical weight), which couple to the 
low-lying states via recombination cascades. 
For the case of \ion{C}{ii} this implies: a) an increased sensitivity of the
\ion{C}{ii/iii} ionization balance to the exact run of the photoionization
cross-sections of levels at low excitation energies and b) an increased
sensitivity of transitions like \ion{C}{ii} $\lambda\lambda$6151, 
6462 and 4267\,{\AA} to non-LTE effects because of their
participation in the recombination~cascade.

\begin{figure*}[ht!]
\centering
\includegraphics[width=.8\linewidth]{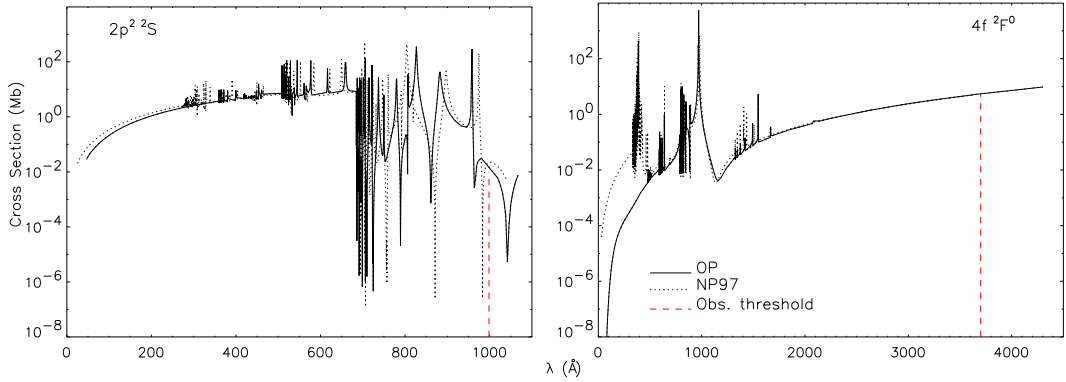}
\caption{Comparison of \ion{C}{ii} photoionization cross-sections from the Opacity Project (Yan \& 
Sea\-ton~\cite{yansea87}) and Nahar~(\cite{nahar97}) for 2p$^2$~$^2$S and 4f~$^2$F$^{\rm o}$
as a function of wavelength. See the text for a discussion.}
\label{RBFlevel}
\end{figure*}

\begin{figure}[ht!]
\centering
\includegraphics[width=0.72\linewidth]{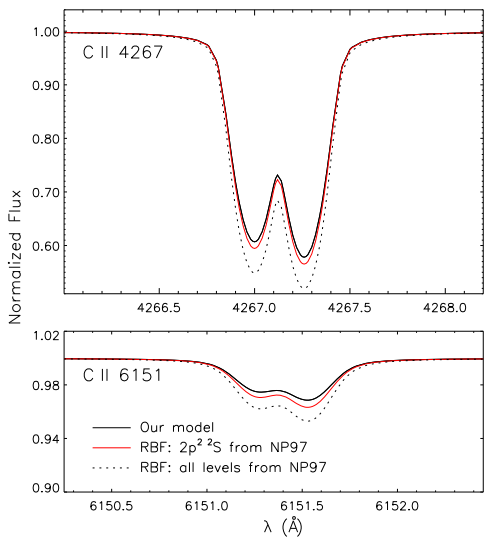}
\caption{Sensitivity of line profiles of two \ion{C}{ii} transitions to variations of 
photoionization cross-sections. The calculations are made
for three different model atoms using the same model atmosphere (as
appropriate for HR\,5285): our reference model atom,
the reference model atom with cross-sections for 
2p$^{2}~^{2}$S from NP97,
and the reference model atom with
all photoionization data replaced by values from NP97. 
The profiles are not convolved for effects of rotation or instrumental broadening.}
\label{RBFlines}
\end{figure}

A comparison of total photoionization cross-sections from OP 
(Yan \& Seaton~\cite{yansea87}) and NP97
for two levels (marked in the Grotrian diagram, Fig.~\ref{grotrian}) is given
in Fig.~\ref{RBFlevel}. The first term, 2p$^2$~$^2$S, is not
directly involved in the formation of the \ion{C}{ii} $\lambda\lambda$\,4267, 6151 and
6462\,{\AA} transitions. However, it is populated
considerably and therefore contributes to the \ion{C}{ii/iii} ionization
balance. The cross-section for photoionization from this level to the ground state of
\ion{C}{iii} is small because of radiative selection rules. Shortward of
$\sim$700\,{\AA} the cross-section rises by about two orders of magnitude as
the optically allowed channel of photoionization to the first excited level
of \ion{C}{iii} opens. The overestimation of the photoionization rate 
from the 2p$^2$~$^2$S state by adopting the total instead of partial 
cross-sections is however insignificant. The contribution of wavelengths
below $\sim$700\,{\AA} to $R_{ij}$ is e.g. less than 1\% 
in the example discussed below.
Note the wavelength shifts in the resonance structures of the OP and the NP97 data.
This results in different contributions of the region between the threshold and the
Lyman edge 
to the integral in $R_{ij}$, which affect the photoionization rates 
considerably. On the other hand,
the photoionization cross-sections for 4f~$^2$F$^0$ agree well, except for
the resonance structure at shortest wavelengths where the stellar flux becomes 
negligible. Consequently, both data (and many others for highly-excited
levels) are exchangeable without showing consequences 
for the spectrum synthesis computations. Note that we adopt experimental threshold 
wavelengths.
On the other hand, for 2p$^2$~$^2$S and other levels at lower excitation energy 
there is a non-negligible effect on the $\lambda\lambda$4267 and 6151\,{\AA}
transitions when exchanging both data, as can be seen in Fig.~\ref{RBFlines}.
In an extreme case, accounting for photoionization cross-sections from NP97 for all
levels results in a very strong \ion{C}{ii} $\lambda\lambda$4267\,{\AA} line.
A reduction of the C abundance by up to $\sim$0.8\,dex is required to
fit the observed line profile in our calibration stars with such a model atom.
As a consequence of this we give preference to the OP data over the cross-sections of
NP97 in our final model atom (`model of reference'), 
which helps us to reproduce observation over the entire parameter range 
consistently.\\ [-2.5mm]

\begin{figure}
\centering
\includegraphics[width=.85\linewidth]{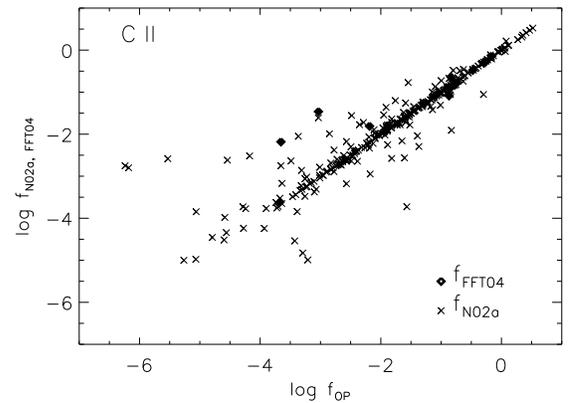}
\caption{Comparison of oscillator strengths from Nahar~(\cite{nahar02a}: N02a) and Froese Fischer \& Tachiev 
(\cite{FFT04}: FFT04) vs. values from the Opacity Project (Yan et al.~\cite{yan87}: OP).}
\label{fvalues}
\end{figure}
\emph{Oscillator strengths.} 
Comparisons between multiplet $f$-values from three sources of {\em ab-initio} 
computations are shown in Fig.~\ref{fvalues}:
data based on {\sc i}) the multiconfiguration Hartree-Fock method 
in the Breit-Pauli approximation of FFT04, 
{\sc ii}) the Breit-Pauli {\bf {\emph R}}-matrix method of N02a and {\sc iii}) results
from {\bf {\emph R}}-matrix calculations assuming LS-coupling, as obtained by the Opacity 
Project (Yan et al.~\cite{yan87}). The primary source of $f$-values is FFT04,
which should be most accurate. Our preference for the OP over the N02a data
is motivated by the good agreement of the former with FFT04 (with two
exceptions), while oscillator strengths from N02a may show large differences for several lines. 
Data from N02a is therefore adopted only in the cases where the other sources do not provide 
information.\\ [-2.5mm]

\begin{figure}[!t]
\centering
\includegraphics[width=0.72\linewidth]{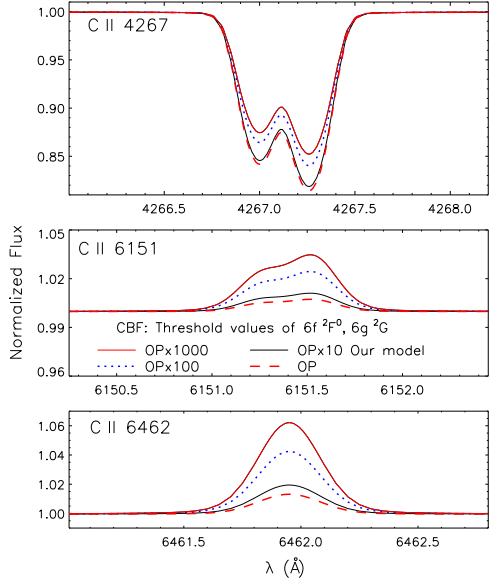}\\[2mm]
\caption{Reactions of three highly non-LTE-sensitive lines 
to changes of collisional ionization cross-sections. 
The modifications are made for two energy levels which are directly involved in 
the formation of \ion{C}{ii}~$\lambda\lambda$6151 and 6462\,{\AA} and indirectly in \ion{C}{ii} 
$\lambda$4267{\AA}. % (see Grotrian diagram). 
The calculations are made for different model atoms with specific values of the reaction 
cross-section at threshold 
and the same set of atmospheric parameters (as appropriate for $\tau$~Sco).} 
\label{CBF}
\end{figure}

\emph{Collisional ionization cross-sections.} 
Details of the colli\-sional ionization cross-sections are 
important mostly for high-excitation levels ($n$$>$$5$). Only for these a 
 significant amount of electrons in the Maxwell distribution 
are energetic enough to overcome the threshold for the reaction.
We use the approximation of Seaton~(\cite{seaton62}), which is expected to
be accurate to an order of magnitude at best, to evaluate the collisional 
ionization rates because of a lack of any data from {\em ab-initio} computations.
Good agreement between the modelling and most observed \ion{C}{ii} lines is
achieved (see below) when using the standard recipe of applying threshold
photoionization cross-sections, except for two observed transitions.
Improvements for the \ion{C}{ii} $\lambda\lambda$6151 and 6462\,{\AA} 
lines may be obtained by an empirical scaling of threshold cross-sections for the
6f~$^2$F$^o$ and 6g~$^2$G levels, the upper levels of the transitions.
An increase by a factor~10, i.e. within the expected uncertainties,
helps to reproduce observation over the entire atmospheric parameter range.
The effects of scaled collision rates for those line profiles
are shown in Fig.~\ref{CBF}, as well as \ion{C}{ii} $\lambda$4267\,{\AA}, 
which remains practically unaffected.
Larger scaling factors are empirically excluded. 
\\ [-2.5mm]

\begin{figure}[ht!]
\centering
\includegraphics[width=0.7\linewidth]{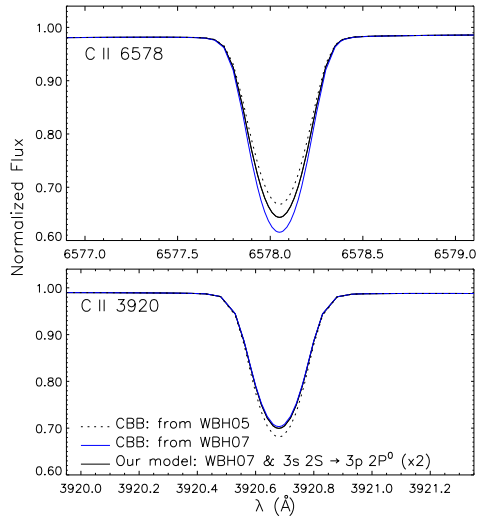}\\
\caption{Effect of employing different effective collision strengths on the 
\ion{C}{ii} $\lambda\lambda$ 6578 and 3920 \AA~lines (for HR\,1861). Only
variations of the data from {\em ab-initio} calculations (Wilson et
al.~\cite{WBH05}, \cite{WBH07}) are considered.}
\label{CBB}
\end{figure}
\begin{figure}
\centering
\includegraphics[width=.94\linewidth]{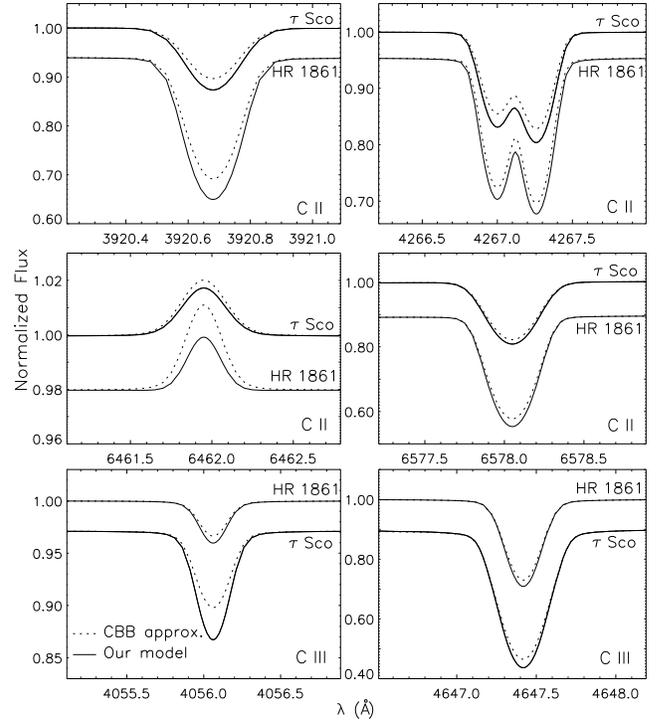}
\caption{Comparison of synthetic \ion{C}{ii/iii} line profiles for two of our sample
stars using different model atoms. One calculation uses our reference model atom and 
the other adopts standard approximation formulae for evaluating collisional
rates for all transitions.}
\label{CBB2}
\end{figure}
\begin{figure}[t!]
\centering
\includegraphics[width=.9\linewidth]{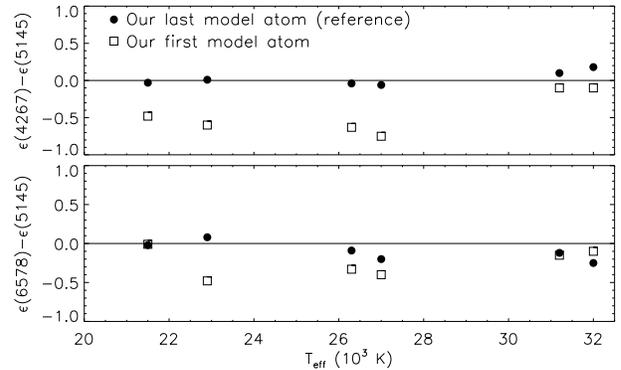}
\caption{ Abundance differences from the analysis of 
\ion{C}{ii}\,$\lambda\lambda$4267 and 6578\,{\AA} (highly
sensitive to the input atomic
 data) and 
\ion{C}{ii}\,$\lambda$5145\,{\AA} ('in LTE').
 Displayed are results for our six sample stars, as a function of effective temperature.
A comparison of our initial model atom (all radiative transition data for 
 \ion{C}{ii} from N02a, NP97) and our final model atom after the calibration 
is shown. See the text for details.}
\label{oldnew}
\end{figure}
\emph{Collisional excitation cross-sections.}
The collisional data used in a model atom will have an influence on the
predicted line profiles.
Accurate data from {\em ab-initio} calculations for larger sets of transitions 
have become available only recently. We employed effective collision strengths of
Wilson et al.~(\cite{WBH05}) to construct the \ion{C}{ii} model. A later
revision of part of the data (Wilson et al.~\cite{WBH07}) had negligible 
influence on the predicted line profiles of almost all observable
transitions, except for \ion{C}{ii} $\lambda\lambda$\,3918/20 and
6578/82\,{\AA}, see Fig.~\ref{CBB} for an example. The good agreement of
abundances derived from these four with other transitions was broken when
using the improved Wilson et al.~(\cite{WBH07}) data, requiring
abundance adjustments of up to $\sim$0.3\,dex to match observation. We find
that the situation may be improved for the stars of our calibration sample 
by increasing the effective collision
strength for the 3s\,$^2$S--3p\,$^2$P$^{\rm o}$ transition by an empirical
factor of two, see Fig.~\ref{CBB}. This is larger than the typical
uncertainty of such {\em ab-initio} data, which amounts to an estimated 10-20\%.
However, a closer inspection of the energy-dependent collision strength for
this transition shows that resonances dominate $\Omega$ in the region near threshold.
The positions and strengths of the resonances are sensitive to the details
of the atomic data calculations, in particular to the assumptions made
 for constructing the target. 
Consequently, more comprehensive
{\em ab-initio} computations are required to investigate this in detail.
However, these are beyond the scope of the present work.

Collisional data from {\em ab-initio} calculations are typically
available only for transitions between relatively low-lying energy levels.
In the case of \ion{C}{ii} the dataset is complete for levels up to
principal quantum number $n=4$. Therefore, for the bulk of the transitions
approximation formulae have to be applied. However, trends and regularities 
from the {\em ab-initio} data may be used to improve on the standard
approximations made for these, e.g. dropping the assumption of an 
energy-independent $\Omega=1$ for the evaluation of the Allen (\cite{allen73}) 
formula (see Sect.~\ref{atom}).
In practice, simple approximations are used to describe collisional processes
in most of the model atoms available for non-LTE calculations at present.
One can ask what the effects on synthetic profiles will be from such a
simplification. This is shown in Fig.~\ref{CBB2} for selected lines in two
of our programme stars. Notable corrections in abundance would be required~to
reproduce the results from our reference model, an increase in some of
the lines but also a decrease in others. Our overall good agreement with observation
(Sect.~\ref{sect_results}) would be destroyed: the non-LTE sensitive lines
and the 'lines in LTE', which are unaffected by such modifications, would
indicate widely different abundances. Note that the effects vary from star
to star.

\begin{figure*}
\centering 
\includegraphics[width=.72\linewidth]{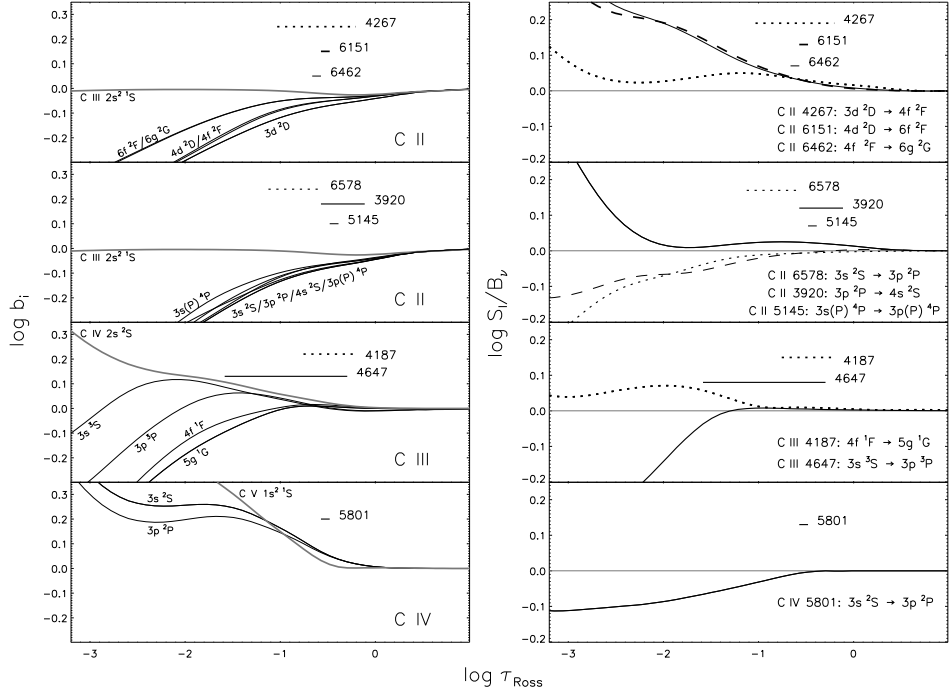}
\caption{ Departure coefficients $b_i$ and ratio of line source-function $S_\mathrm{l}$ 
to Planck function $B_\nu$ at line centre as a function of $\tau_\mathrm{Ross}$ for 
selected carbon lines in $\tau$~Sco. 
The spectral lines are encoded by the different line styles indicating the
 line-formation depths.
Thin grey lines on the right-hand panel correspond to $S_\mathrm{l}$= $B_\nu$.
}
\label{lineformation}
\end{figure*} 

We conclude the discussion on the impact of atomic data on line-formation 
calculations in Fig.~\ref{oldnew}. Here, we quantify the differences in
abundance derived with our initial model atom (built from the available homogeneous
set of atomic data, NP97/N02a) and with our model of reference (after the 
calibration) for the entire sample of stars. Abundances from 
the strong and non-LTE-sensitive \ion{C}{ii} $\lambda\lambda$4267 and 6578\,{\AA} 
transitions are compared to abundances derived from the weaker
\ion{C}{ii} $\lambda$5145\,{\AA} line, which is almost insensitive to the
details of the model atom (i.e. it is `in LTE'). The relative abundance is
displayed as a function of effective temperature for each star
(see Section~\ref{sect_results} for final results). Pronounced systematic
trends exist when the initial model atom is used, correlating with the strength 
of the lines (i.e. stronger lines show a larger sensitivity to non-LTE effects 
and therefore to the input atomic data). These trends and abundance
differences from \ion{C}{ii} $\lambda\lambda$4267 and 6578\,{\AA} almost
vanish when our model of reference is used. The remaining small differences
may be reduced even further when improved atomic data, in particular for
collisions involving high-excitation levels, become available. Note, that
the \ion{C}{ii} $\lambda$6582\,{\AA} line (not displayed here) shows a trend 
similar to that of $\lambda$6578\,{\AA}. The whole multiplet centred on \ion{C}{ii} 
$\lambda$5145\,{\AA} also behaves consistently.
This kind of test has been made for all \ion{C}{ii-iv} lines for every set of
input atomic data in the empirical calibration process of the model~atom.
The detailed study of non-LTE and LTE abundances in Sect.~\ref{sect_results} helps to identify
the lines in LTE for the different atmospheric parameters under analysis.
These may provide good starting points for further analyses when one desires
to avoid non-LTE effects.\\[-5.5mm]

\subsection {Line-formation details}
A closer study of the underlying line-formation processes allows 
the nature of the non-LTE effects to be understood. 
This is shown in Fig.~\ref{lineformation} for some representative 
transitions of \ion{C}{ii/iii/iv} in our programme star $\tau$~Sco.
Departure coefficients $b_i=n_i^\mathrm{NLTE}/n_i^\mathrm{LTE}$ 
are displayed in the left panel of Fig.~\ref{lineformation}
for the levels involved in the transitions of interest and the ion ground states. 
Non-LTE 
departures of the level occupations impact the line source function
$S_\mathrm{l}$. We recall that the ratio of $S_\mathrm{l}$ to the Planck
function $B_\nu$ (shown for selected transitions in the right panel of Fig.~\ref{lineformation}) is
 $ S_\mathrm{l}/B_{\nu} = [\exp{(h \nu_{ij} /kT)} - 1] /[b_i/b_j
\exp(h \nu_{ij} /kT) - 1]$. 
$S_\mathrm{l}/B_{\nu}$ is determined
by the ratio of the departure coefficients for the lower and upper levels
($i$, $j$). An overpopulation of the upper level relative to the lower (i.e.
$S_\mathrm{l}/B_{\nu}>1$) results in non-LTE weakening of the line and may
lead to emission in cases of pronounced overpopulation, 
while the inverse gives non-LTE strengthening.

The level populations reach detailed equilibrium values deep in the
atmosphere, where large collisional rates and small mean-free paths between
photon absorptions (both because of the high densities of the plasma)
enforce this inner boundary condition.
Double-ionized carbon is the main ionization stage at the temperatures of 
$\tau$\,Sco - the \ion{C}{iii} ground state is close to LTE. Single-ionized
carbon is overionized at line-formation depths, therefore the levels 
are underpopulated relative to LTE, and \ion{C}{iv} and the \ion{C}{v} 
ground state are overpopulated.

In general, level populations in \ion{C}{ii} depart most from detailed
equilibrium in the low-excitation states and approach LTE values gradually
with increasing excitation energy, as collisions facilitate coupling with
the \ion{C}{iii} ground state. Therefore, most of the non-LTE-sensitive transitions 
in \ion{C}{ii} have upper levels that are overpopulated relative to the
lower level, such that the lines experience slight ($\lambda$\,3920\,{\AA})
to notable weakening ($\lambda$\,4267\,{\AA}) relative to LTE and may even turn into
emission ($\lambda\lambda$\,6151, 6462\,{\AA}). The
$\lambda\lambda$\,6578/82\,{\AA} doublet experiences non-LTE 
strengthening for lower effective temperatures, while in hotter objects like
$\tau$\,Sco the lines are found to be close to LTE. The other observable
\ion{C}{ii} lines arise in the quartet spin system. They are weaker 
(\ion{C}{ii} $\lambda$\,5145\,{\AA} is the strongest), i.e.
they are formed deeper in the atmosphere, and their formation involves
high-excitation levels which are coupled collisionally at these depths, 
such that these lines are essentially in LTE. Note that the behaviour of
$S_\mathrm{l}/B_{\nu}$ for $\lambda$\,4267\,{\AA} is in good agreement with the
findings of Sigut~(\cite{Sigut96}). On the other hand, notable differences exist for 
the $\lambda\lambda$\,6578/82\,{\AA} doublet in particular at higher
$T_\mathrm{eff}$. The reasons for this will be discussed in the next section
(but see also Fig.~\ref{EWvsT}).

The \ion{C}{iii} transitions can also experience both, non-LTE weakening (like the
strong $\lambda$\,4187\,{\AA}) and non-LTE strengthening (like the strong
triplet 4647-4651\,{\AA}). The \ion{C}{iv} doublet
$\lambda\lambda$\,5801/12, which becomes observable only in our hottest
stars, shows a pronounced non-LTE strengthening.

\subsection {Sensitivity of $\varepsilon$(C) to atmospheric parameter variations}\label{param_variations}

Spectral lines of carbon, like many other metal lines, can react sensitively to 
variations of the stellar atmospheric parameters. This property provides us
with a powerful tool for the atmospheric parameter and abundance determination,
using our iteration scheme. This is a non-trivial and crucial step in the
analysis, which must be performed carefully in order to avoid systematic error.
We investigate consequences of systematically biased atmospheric parameters
on carbon line profiles and the derived abundances in the following
\footnote{Note that an underestimated source of systematic error in view of line blanketing are abundance `standards' in the the widely used classical Kurucz~(\cite{kur93a}) ODFs, computed 
on the basis of (scaled) solar abundances according to Anders \& Grevesse~(\cite{AnGr89}).  
Employing ODFs with appropriately reduced metallicity for the
model atmosphere calculations (see Paper~1 for a discussion) decreases
the atmospheric temperature structure via a backwarming
effect, by up to $\sim$500\,K in the line-formation region, i.e. slightly higher than our uncertainties in $T_\mathrm{eff}$.}.

\begin{figure}[t!]
\centering
\includegraphics[width=0.85\linewidth]{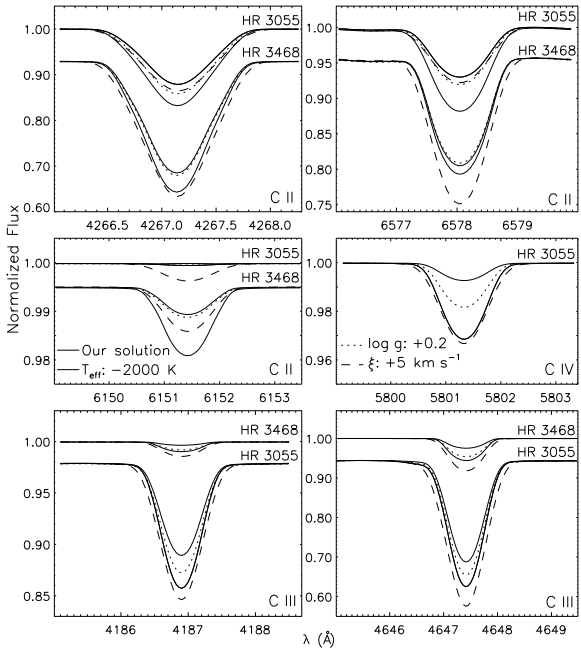}
\caption{ Sensitivity of selected \ion{C}{ii/iii/iv} lines to 
atmospheric parameter variations in two giants: 
HR\,3055 (hotter, B0\,III) and HR\,3468~(cooler, B1.5\,III). Our 
solution corresponds to the final parameters from Table~\ref{parameters} and 
a constant value of $\varepsilon(\rm{C})$ for all lines. The parameter offsets
are typical for statistical and systematic uncertainties
from published values. The theoretical spectra are convolved
with a rotational profile for $v\sin i= 20~\mathrm{km~s^{-1}}$. 
Our solution establishes the ionization equilibrium.
The star HR\,3468 is too cool to show \ion{C}{iv} lines.
}
\label{sens_profiles}
\end{figure}

\begin{table}[ht!]
\caption[]{Systematic uncertainties in carbon abundances (in dex, relative to our final
results, Table~3) caused by atmospheric parameter variations and the assumption of
LTE for the line-formation calculations.} 

\label{param_abund}
\vspace{-4mm}
 $$
 \begin{array}{lccccc}
 \noalign{}
  \hline
  & \mathrm{HR\,3055}& \Delta T_\mathrm{eff} & \Delta \log\,g & \Delta\,\xi
  & \mathrm{LTE}\\
   &   & -2000\,{\rm K}        & +0.2\,{\rm dex}& +5\,{\rm km\,s}^{-1}&\\
 \hline     
\ion{C}{ii}&4267.2  &-0.33 &-0.11 &-0.16  &-0.40  \\
           &5133.3  &-0.30 &-0.10 &~~0.00 &~~0.00  \\
           &5143.4  &-0.40 &-0.05 &~~0.00 &~~0.00  \\ 
           &5145.2  &-0.32 &-0.09 &-0.02  &~~0.00  \\ 
           &5151.1  &-0.30 &-0.08 &~~0.00 &~~0.00  \\
           &5662.5  &-0.33 &-0.13 &~~0.00 &~~0.00  \\ 
           &6578.0  &-0.40 &-0.15 &-0.10  &-0.01  \\ 
           &6582.9  &-0.30 &+0.02 &+0.05  &+0.03  \\[1mm] 
\ion{C}{iii} &4056.1 &+0.21 &+0.06 &-0.04  &+0.08   \\
             &4162.9 &+0.28 &+0.09 &-0.03  &+0.25   \\ 
             &4186.9 &+0.35 &+0.15 &-0.08  &+0.07   \\
             &4663.5 &+0.22 &+0.07 &-0.03  &+0.22   \\ 
             &4665.9 &+0.26 &+0.08 &-0.08  &+0.35   \\ 
             &5272.5 &+0.16 &+0.01 &~~0.00 &~~0.00   \\[1mm]
\ion{C}{iv}&5801.3 & +1.06 &+0.46& -0.03  &+0.39\\ 
           &5811.9 & +1.06 &+0.46& -0.03  &+0.39   \\
 \hline
 \end{array}
 $$
 \end{table}

The offsets for the parameters (in effective temperature 
$\Delta$\,$T_\mathrm{eff}$= $-$2000\,K, surface gravity
$\Delta$\,$\log g$=\,$+$0.2\,dex and microturbulent velocity
$\Delta$\,$\xi$=\,$+$5\,km\,s$^{-1}$) are
representative for systematic discrepancies between our final values and
those from previous studies (and also among previous studies). Note that they are much larger than
our statistical uncertainties. Such discrepancies may be caused by several 
factors, among others: 
{\sc i}) photometric effective temperature calibrations based on model
atmospheres with insufficient line blanketing,
{\sc ii}) spectroscopic ionization equilibria based on predictions of
incomplete model atoms, 
{\sc iii}) the assumption of LTE for the computation of Balmer line profiles, which may cause 
a systematic overestimate of $\log g$ by up to 0.2\,dex (see Paper~1).

Figure~\ref{sens_profiles} shows the sensitivity of selected \ion{C}{ii/iii/iv} lines 
in two giants of our sample to variations of like $T_\mathrm{eff}$, $\log g$ and $\xi$.
The profiles accounting for these variations are 
compared with those computed with our final atmospheric 
parameters and an averaged carbon abundance of $\varepsilon(\rm{C})$\,$=$\,8.32. 
The sensitivity to parameter variations differs from line to
line.

For the hotter stars, the \ion{C}{iv} and -- when strong enough -- the \ion{C}{ii} 
$\lambda$\,6151\,{\AA} multiplets are ideal indicators for both 
$T_\mathrm{eff}$ and $\log g$, while the rest of the lines are mostly 
sensitive to changes in $T_\mathrm{eff}$ and the strong \ion{C}{ii/iii} lines
also to changes in $\xi$. 
For the cooler stars, the \ion{C}{ii} $\lambda\lambda$\,4267 and 6151\,{\AA} multiplets are highly sensitive to variations of $T_\mathrm{eff}$ and $\xi$,  
the rest of the lines are mostly sensitive to $T_\mathrm{eff}$ and 
the strong lines also to $\xi$.

\begin{figure*}[ht!]
\sidecaption
\includegraphics[width=7.5cm, angle=90]{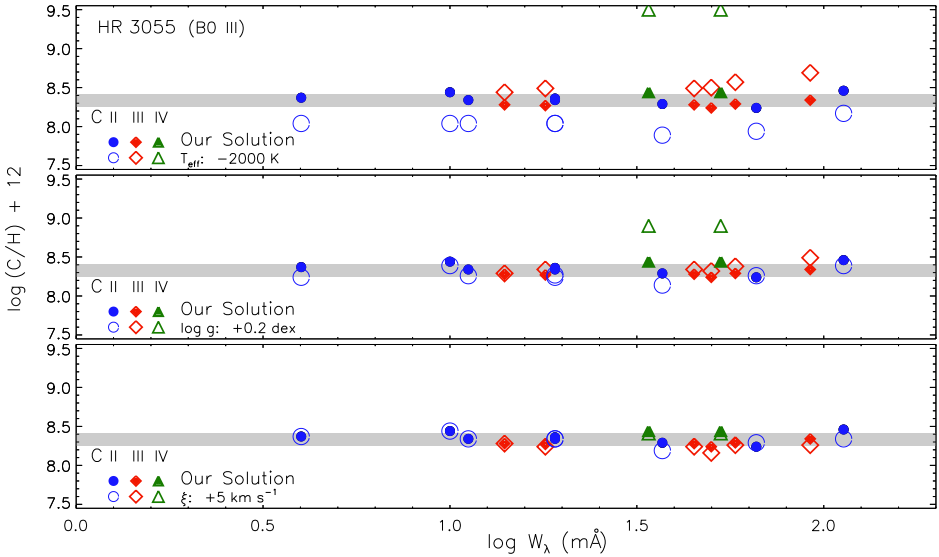}
\caption{ Sensitivity of carbon abundances to stellar parameter
variations: $T_\mathrm{eff}$ (upper panel), $\log~g$ (mid panel) and
microturbulent velocity (lower panel).
The offsets of the parameters are displayed in the lower left part of each panel.
A large spread in abundance 
from individual lines of the three ionization stages results, in particular 
for variations of $T_\mathrm{eff}$ (upper panel).
Note also the 
implications of using only few lines of one ionization stage for abundance
determinations in the presence of systematic errors in the atmospheric
parameters.
The grey bands correspond to 1$\sigma$-uncertainties of the stellar carbon 
abundance in our solution (summarised in
Table~\ref{parameters}).\vspace{5.5mm}}  
\label{sens_abund}
\end{figure*}

The response to variations in $\log g$ can be amplified for \ion{C}{ii}
$\lambda\lambda$\,6578/82\,{\AA} at higher gravities, because they are
formed on the red wing of H$\alpha$. The local continuum and
therefore the line-formation depths may change as a consequence, 
impacting the non-LTE effects. A correct treatment of the hydrogen Balmer
line-opacity therefore plays an important r\^ole in this context. Non-LTE effects 
strengthen the Balmer line wings in particular at higher temperatures
(Paper~1).

For HR~3055 we have quantified the systematic effects exemplified in
Fig.~\ref{sens_profiles} (note that \ion{C}{iii} $\lambda$4647\,{\AA} is
blended and \ion{C}{ii} $\lambda$6151\,{\AA} is too weak in this case)
by deriving non-LTE carbon abundances for the modified values of 
$T_\mathrm{eff}$, $\log g$ and $\xi$ and comparing them to our solution for 
individual lines. This is summarised in Table~\ref{param_abund} and visualised 
in Fig.~\ref{sens_abund}. Table~\ref{param_abund} also shows systematic
offsets that arise from the assumption of LTE for the line-formation
calculations. This comparison allows us to identify the relative importance
of atmospheric parameters/non-LTE effects for some key spectral lines.
%(often termed `non-LTE corrections'). 
Note that the \ion{C}{iv} lines are extremely sensitive to changes 
in $T_\mathrm{eff}$ and $\log g$ 
at this temperature (31\,200~K), with discrepancies amounting to up to $\sim$$+$1.0\,dex 
in abundance for $\Delta$\,$T_\mathrm{eff}$= $-$2000\,K\footnote{Even larger
offsets in $T_\mathrm{eff}$ are found with respect to the literature, up to 
$\Delta T_\mathrm{eff}\simeq-4000$\,K, see Section 5.}. 
The \ion{C}{ii/iii}
ionization balance is also never established for a variation of these 
parameters (abundance changes
down to $-$0.40\,dex for \ion{C}{ii} and up to $+$0.35\,dex for \ion{C}{iii}
when compared to our solution).
An expected reduction of the abundances from strong lines is obtained 
for an increased microturbulence. 
Note that the systematic variations of carbon abundance with $\xi$
for some lines are significant considering the high accuracy we are aiming
at, despite smaller effects in general than for $T_\mathrm{eff}$ and $\log
g$ variations. 
 
The solutions for the modified atmospheric parameters are characterised by 
a large scatter of C abundances from indivi\-dual lines (the statistical 1-$\sigma$ uncertainties 
increase by up to $\sim$0.4 dex). 
Note that variations of $T_\mathrm{eff}$ show by far the largest effects.

We conclude that
the use of only a few spectral lines from one ionization stage for C abundance
determinations -- which is common practice in the literature -- may have serious
implications on the final accuracy. Possible systematic discrepancies, as
indicated here, may remain unrecognised. Moreover, the opportunity to improve on the 
atmospheric parameter determination by establishing the highly
parameter-sensitive ionization balance may be missed in such cases.
This high sensitivity of the carbon lines to atmospheric parameter variations
can be 
used as an important tool for precise quantitative spectral analyses of this kind
of star when applying the calibrated model atom in the future.

Some lines are practically unaffected by non-LTE effects, as indicated by
Table~\ref{param_abund} (see also Sect.~\ref{atomic_sensitivity}).
This implies that they are almost insensitive to any reasonable choice of
model atom. On the other hand,
they are highly sensitive to the choice of atmospheric parameters. 
This property is highly useful for the model atom calibration because it helps to
disentangle effects due to non-LTE from those due to inaccuracies in the
stellar parameters, facilitating a reduction of systematic errors.
Typical systematic
uncertainties in the atmospheric parameters can have a similar or even larger
impact on the carbon abundance determination than a neglect of non-LTE
effects.

\begin{figure*}[ht!]
\centering
\sidecaption
\includegraphics[width=12cm]{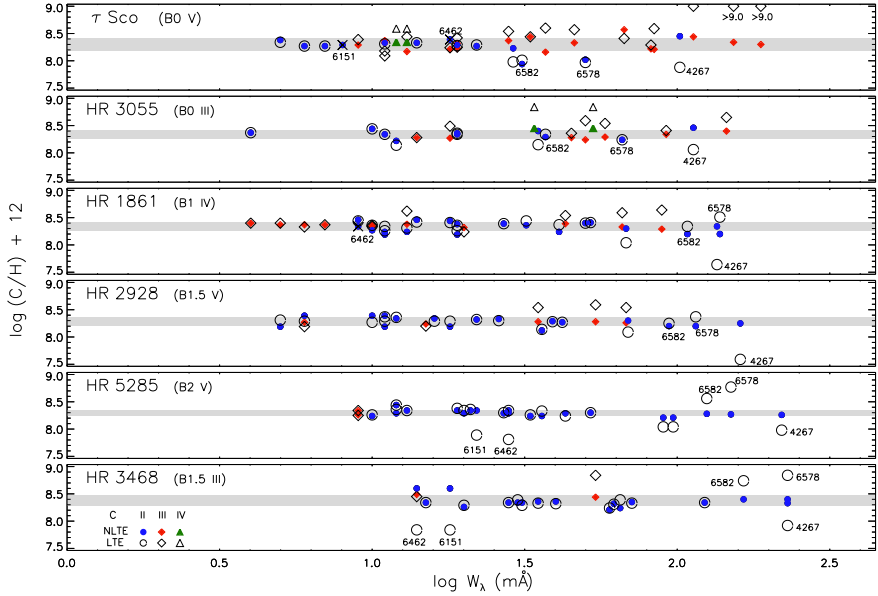}
\caption{ Non-LTE and LTE abundances derived from line profile fits to individual
  \ion{C}{ii-iv} lines in the programme stars as a function of equivalent width.
The ID of the stars and the spectral classification is given in the upper 
left corner of each panel.
The grey rectangles correspond to 1$\sigma$-uncertainties 
of the stellar carbon abundance from the
line-to-line scatter. Identification of lines with high sensitivity to 
non-LTE effects is displayed. Emission lines are marked by crosses (\ion{C}{ii} 
$\lambda\lambda$6151 and 6462\,\AA~in $\tau$\,Sco and $\lambda$6462\,\AA~in HR\,1861): 
LTE calculations are not able to reproduce them even qualitatively.
}
\label{results}
\end{figure*}

\begin{table*}[t]
\caption[]{
Stellar parameters and carbon abundances of the programme stars.\\[-5mm]}
\label{parameters}
\vspace{-5mm}
$$
\begin{array}{lccccccr}
\noalign{\smallskip}
 \hline\
&\mathrm{\tau\,Sco}& \mathrm{HR\,3055}& \mathrm{HR\,1861}& \mathrm{HR\,2928}& \mathrm{HR\,5285}& \mathrm{HR\,3468} \\%[1mm]
&\mathrm{HD\,149438}&\mathrm{HD\,63922}&\mathrm{HD\,36591}&\mathrm{HD\,61068}&\mathrm{HD\,122980}&\mathrm{HD\,74575}\\
\hline\\[-3mm]
           & \mathrm{Sco\,Cen} &  \mathrm{Field} &  \mathrm{Ori\,OB1} &
	   \mathrm{Field} &  \mathrm{Sco\,Cen} &  \mathrm{Field}\\[.5mm]
T_\mathrm{eff}~(\mathrm{K})      &32000\pm300  &31200\pm300    &27000\pm400
&26300\pm400   &21500\pm400  &22900\pm400\\[.5mm]
\log g\,\mathrm{(cgs)}     &4.30\pm0.05       &3.95\pm0.05
&4.12\pm0.05       & 4.15\pm0.05       &4.20\pm0.05       &3.60\pm0.05\\[.5mm]
\xi\,\mathrm{(km\,s^{-1})}   &5\pm1      &  8\pm1   & 3\pm1    & 3\pm1
&4\pm1     & 5\pm1 \\[.5mm]
v\sin i\,\mathrm{(km\,s^{-1})}&4\pm2  &  29\pm4   & 12\pm1   & 14\pm1
&18\pm1     & 11\pm2 \\[.5mm]
\mathrm{\zeta\,(km\,s^{-1})} &4\pm2    &  37\pm8  &\ldots        & 20\pm2   &\ldots        & 20\pm1 \\[.5mm]
y\,(\mathrm{by~number}) & 0.089\pm0.01 & 0.080\pm0.01 & 0.089\pm0.01 & 0.089\pm0.01 & 0.089\pm0.01 & 0.089\pm0.01\\[.5mm] 
\mathrm{\varepsilon(\mathrm{C})~(dex)}&8.30\pm0.12 &8.33\pm0.08 &8.33\pm0.08 & 8.27\pm0.07 &8.29\pm0.05 & 8.37\pm0.10  \\[.5mm]
\#\,\mathrm{C~lines}  &  32  &  19  &  30  & 22   &  22   & 19  \\[1mm]
\hline\\[-7mm]
\end{array}
$$
\end{table*}

\section{Results}\label{sect_results}

Accurate atmospheric parameters and carbon abundances are derived from 
line-profile fitting by $\chi^2$ minimisation, which puts tighter
constraints than matching only equivalent widths. These are summarised in 
Table~\ref{parameters} for our six programme stars
(as obtained from our iterative process, Fig.~\ref{hhec}). 
The uncertainties of $T_\mathrm{eff}$ and $\log g$ are estimated
from the extremely sensitive carbon ionization equilibrium. 
They are lower for hotter stars because of the additional restrictions
imposed by the presence of lines from three ionization stages.
For velocities, we provide the 1$\sigma$-uncertainties from the analysis of 
the entire line ensemble. Projected rotational velocity and 
(radial-tangential) macroturbulence (Gray~\cite{Gray92}, p. 407ff.)
were simultaneously derived allowing for small line-to-line variations 
in order to obtain an optimum fit 
(see Ryans et al.~\cite{ryans02} for the case of B supergiants).

We should emphasise that the final set of atmospheric parameters 
as derived from the \ion{C}{ii/iii/iv} ionization balance is in agreement with 
those from our previous quantitative analyses of these stars. In particular,
a simultaneous match is achieved for  
{\sc i}) the H and He lines in the visual and (where available) 
in the near-IR, including the \ion{He}{i/ii}
ionization equilibrium in the hotter stars (Paper~1, also the source of
helium abundance $y$) and
{\sc ii}) the spectral energy distributions from the UV to the near IR (Paper~2). 
A similar degree of consistency is typically not obtained in comparable studies of early-type
stars.

A large quantity of carbon lines is analysed for the first time, giving consistent
abundances for all of them. The excellent quality of the observed
spectra combined with our improved analysis technique allow us to
achieve such precise results. The 1$\sigma$-uncertainties from the
line-to-line scatter are typically of the order 0.05--0.10\,dex. We estimate
the systematic uncertainties due to remaining errors in atmospheric parameters 
and atomic data to be of the order 0.10--0.15\,dex, taking
Table~\ref{param_abund} and our experiences from
Sect.~\ref{atomic_sensitivity} as a guideline.

Details on the analysis of individual lines can be found in
Table~3. This summarises line identifications, level designations, 
$\log gf$ values, excitation potential of the lower level $\chi_l$, 
and for each star equivalent widths $W_{\lambda}$ and the derived non-LTE and LTE 
abundances. 
Note that we analyse as many lines as possible per star,
excluding only features with strong blends by other chemical species.
This helps us to understand the behaviour of each line and the quality of our modelling 
at different temperatures and gravities (see the previous section). 
Transitions involving autoionizing states are not accounted for explicitly 
in the spectrum synthesis 
(however, they are considered via resonances in the photoionization cross-sections).
These are lines like \ion{C}{ii} $\lambda\lambda$\,4075 (a blend with \ion{O}{ii}), 4318
(a blend with \ion{S}{ii}), 4374.3, 4375.1, 4376.6, 4411.2/5 and
4627.4\,{\AA}, which have sometimes been used for abundance determinations in
previous studies.

The high quality of our line fits to the observed spectrum of $\tau$\,Sco is
demonstrated in Fig.~\ref{fits_tsco}, available online, for almost all the analysed transitions. 
Similar information for the other programme stars can be found in
Figs.\ref{fits_3055}--\ref{fits_3468} in the online version. Of central
importance is that the abundances derived from the individual lines show a
small scatter in each sample star. Good fits to individual lines can almost always
be obtained, however this does not imply consistency in the entire analysis.
An example are the abundances from Fig.~\ref{sens_abund}, which were
also derived from high-quality line fits, but which show inconsistencies,
expressed as a large line-to-line scatter, 
nonetheless. Another example from many other studies 
are discrepant lines that are excluded from the
analysis in order to reduce the statistical uncertainties.
The present work improves on this because the underlying physics is solved in a more 
consistent way.

We find a single exception to our overall good line fits: the doublet
\ion{C}{ii} $\lambda\lambda$4267.0/2\,{\AA} in $\tau$\,Sco, where the
fine-structure components are resolved (Fig.~\ref{fits_tsco}).
The synthetic profile is slightly broader than the observed one,
even when neglecting micro- and macroturbulent broadening.
Such a detail cannot be observed in the other stars because of their higher
$v \sin i$.

The data from Table~3 are visualised in Fig.~\ref{results}, where abundances
are shown as a function of $W_{\lambda}$. Excellent
consistency is found in the case of the non-LTE analysis
while the quality of the results is considerably degraded in LTE.
Note that the sub-set of the strong lines
provides non-LTE abundances that are in good agreement with those derived from the
entire sample of lines. This proves that our reference model atom is suited
well for analyses of low-S/N spectra and fast-rotating stars, where only the
strongest lines are measurable and where such a consistency check is not
feasible. Note also that the non-LTE effects differ from 
line to line and also from star to star. 

As the carbon abundances presented here are derived 
via line-profile fitting of our synthetic spectra, they do not depend on 
equivalent width measurements. However, careful determinations of $W_\lambda$
via direct integration, cross-checked by Voigt-profile fitting, were performed
to allow a visualisation like in Fig.~\ref{results}.

\begin{landscape}
{\bf Table 3.} 
  \footnotesize
Carbon abundance analysis for the programme stars.\\[2mm] % from individual lines.\\[2mm]
 \label{bigtable}
 \setlength{\tabcolsep}{.15cm}
 \begin{tabular}{lrcrcccccccr}
 \noalign{}
 \hline
        &  &   &   & &$\tau$\,Sco  & HR\,3055   & HR\,1861   & HR\,2928   & HR\,5285 & HR\,3468 \\
Ion 	& $\lambda$\,({\AA}) & l~$-$~u&  $\log~gf$ & $\chi_l$~(eV) &$W_\lambda$~~NLTE~~LTE&
	$W_\lambda$~~NLTE~~LTE&$W_\lambda$~~NLTE~~LTE&$W_\lambda$~~NLTE~~LTE&$W_\lambda$~~NLTE~~LTE &$W_\lambda$~~NLTE~~LTE\\%[1mm]    
\hline
\ion{C}{ii}&  3919.0 &$3p\,^{2}P ~-~ 4s\,^{2}S$ &$-$0.53$^B~$ & 16.33          &    bl    & bl   & bl             & bl& ~90~~8.21~~8.04 &bl          \\
 &3920.6 &      "                               &$-$0.23$^B~$ & 16.33       & ~29~~8.23~~7.98 & ~35~~8.40~~8.15  & ~68~~8.30~~8.04 & ~69~~8.30~~8.09& ~97~~8.21~~8.04 &123~~8.34~~8.34\\
  &4267.0/2 &$3d\,^{2}D ~-~ 4f\,^{2}F^{0}$      &0.56/0.74$^{C+}$&18.04       & 102~~8.45~~7.88 & 113~~8.46~~8.06    & 135~~8.34~~7.64 &161~~8.25~~7.59 & 220~~8.26~~7.98 &230~~8.33~~7.92\\
  &5133.0/3 &$(^{3}P^{0})3s\,^{4}P^{0}~-~ (^{3}P^{0})3p\,^{4}P$&$-$0.21/$-$0.18$^B~$&20.70&~19~~8.29~~8.28 & ~19~~8.34~~8.34 & ~52~~8.41~~8.41 &~42~~8.27~~8.27 & ~52~~8.30~~8.30&~71~~8.35~~8.33\\
 &5137.3 &      "                               &$-$0.91$^B~$&20.70       &        ...        &      ...         &~~9~~8.46~~8.44 & ~~6~~8.39~~8.29& ~12~~8.44~~8.44 &...   \\
 &5139.2 &      "                               &$-$0.71$^B~$&20.70       & ~~5~~8.38~~8.34 &      ...           &~14~~8.46~~8.42 & ~10~~8.39~~8.27& ~13~~8.34~~8.34 &~15~~8.34~~8.34\\
 &5143.4 &      "                               &$-$0.22$^B~$&20.70       & ~11~~8.33~~8.31 & ~10~~8.44~~8.44 & ~27~~8.39~~8.39 & ~26~~8.33~~8.30& ~28~~8.34~~8.34 &~35~~8.36~~8.33\\
 &5145.2 &      "                               &~0.19$^B~$ &20.71        & ~22~~8.27~~8.29 & ~19~~8.36~~8.36 & ~50~~8.40~~8.40 & ~39~~8.29~~8.28& ~43~~8.29~~8.24 &~62~~8.32~~8.31\\
 &5151.1 &      "                               &$-$0.18$^B~$&20.71       & ~14~~8.33~~8.33 & ~11~~8.34~~8.34 & ~32~~8.36~~8.44 & ~22~~8.32~~8.32 & ~27~~8.29~~8.30 &~40~~8.36~~8.32\\
 &5648.1 &$(^{3}P^{0})3s\,^{4}P^{0}~-~ (^{3}P^{0})3p\,^{4}S$&$-$0.42$^B~$&20.70&~~6~~8.27~~8.27 &       ...      &~18~~8.44~~8.41 & ~11~~8.39~~8.37& ~19~~8.34~~8.38 &~28~~8.34~~8.34\\
 &5662.5 &      "                               &$-$0.25$^B~$&20.71        & ~~7~~8.27~~8.27 & ~~4~~8.37~~8.37 & ~19~~8.38~~8.36 & ~16~~8.34~~8.29& ~21~~8.34~~8.36 &~31~~8.34~~8.29\\
  &6578.0 &     $3s\,^{2}S ~-~ 3p\,^{2}P^{0}$   &$-$0.03$^N~$&14.45        & ~50~~8.02~~8.01 & ~66~~8.24~~8.24 & 138~~8.20~~8.51 & 115~~8.20~~8.37& 150~~8.27~~8.77 &230~~8.40~~8.84\\
 &6582.9 &      "                               &$-$0.33$^N~$&14.45        & ~31~~7.94~~7.97 & ~37~~8.29~~8.34 & 108~~8.20~~8.34 &~94~~8.20~~8.25 & 125~~8.28~~8.56 &165~~8.40~~8.74\\
 &6779.9 &$(^{3}P^{0})3s\,^{4}P^{0}~-~ (^{3}P^{0})3p\,^{4}D$&~0.02$^B~$&20.70& ... ~~        &      ...          &~19~~8.19~~8.26 & ~36~~8.12~~8.14& ~33~~8.24~~8.26 &~60~~8.21~~8.24\\
 &6780.6 &       "                              &$-$0.38$^B~$&20.70        & ... ~~            &      ...         &~11~~8.19~~8.26 &  bl          &     bl        & bl         \\
 &6783.1 &       "                              &~0.30$^B~$&20.71         & ... ~~            &~12~~8.22~~8.14 & ~41~~8.24~~8.37 & ~~5~~8.19~~8.31& ~36~~8.24~~8.33 &~65~~8.24~~8.39\\
 &6787.2 &       "                              &$-$0.38$^B~$&20.70         & ... ~~            &      ...          &~11~~8.24~~8.34 & ~11~~8.19~~8.29& ~12~~8.29~~8.36 &~30~~8.34~~8.39\\
 &6791.5 &       "                              &$-$0.27$^B~$& 20.70        & ... ~~            &      ...          &~13~~8.24~~8.31 & ~18~~8.19~~8.29& ~20~~8.29~~8.34 &~20~~8.26~~8.29\\
 &6800.7 &       "                              &$-$0.34$^B~$& 20.71      & ... ~~            &      ...         & ~10~~8.27~~8.36 & ~12~~8.34~~8.36   &~10~~8.24~~8.26 &...&         \\
  &6151.3/5 &$4d\,^{2}D ~-~ 6f\,^{2}F^{0}$      &$-$0.15/0.02$^N~$&20.84  &  $-$8~~8.29~~XX~\,   &      ...          &        ...        & ...         &~22~~8.34~~7.89 &~18~~8.60~~7.84\\
  &6461.9 &$4f\,^{2}F^{0}~-~ 6g\,^{2}G$         &~0.42$^N~$ &  20.95   &  $-$18~~8.39~~XX~~   &      ...          & $-$9~~8.34~~XX~~   & ...           &~28~~8.29~~7.81 &~14~~8.60~~7.84\\[1mm]
 \ion{C}{iii} &4056.1 &$4d\,^{1}D^{0} ~-~ 5f\,^{1}F$&0.27$^B~$&40.20    & ~33~~8.44~~8.44 & ~45~~8.28~~8.36 & ~10~~8.36~~8.36 & ~~6~~8.27~~8.19&   ...            & ...        \\
  &4152.5 &$(^{3}P^{0})3p\,^{3}D~-~ 5f\,^{3}F^{0}$&$-$0.11$^{C+}$&40.06   & ~28~~8.37~~8.54 &      bl    & ~~7~~8.37~~8.37 &      ...         &   ...            & ...        \\
 &4162.9 &       "                          &$-$0.84$^{C+}$& 40.06        & ~46~~8.33~~8.57 &~58~~8.29~~8.54 &         bl     &     bl	         &   ...           & ...      &  \\
  &4186.9 &$4f\,^{1}F ~-~ 5g\,^{1}G$        &~0.92$^B~$&40.01           & ~67~~8.57~~8.41 &92~~8.34~~8.41  & ~20~~8.32~~8.24 & ~15~~8.24~~8.20&   ...            &~14~~8.49~~8.45 \\
  &4515.8 &$4p\,^{3}P ~-~ 5s\,^{3}S$      &$-$0.28$^B~$& 39.40          & ~18~~8.21~~8.24 &               bl  & ~~4~~8.40~~8.40 &      ...	 &   ...             & ...        \\
 &4516.8 &       "                          &$-$0.06$^B~$&39.40         & ~19~~8.23~~8.25 &               bl  & ~~6~~8.37~~8.33 &      ...         &   ...             & ...       & \\
  &4647.4 &$3s\,^{3}S ~-~ 3p\,^{3}P$        &~0.07$^{B+}$& 29.53          & 188~~8.30~~$\ge$9.0&              bl   & ~89~~8.29~~8.64 & ~68~~8.26~~8.54& ~~9~8.26~~8.25 &~54~~8.44~~8.84 \\
 &4650.2 &        "                         &$-$0.15$^{B+}$& 29.53        & 153~~8.34~~$\ge$9.0&               bl  & ~66~~8.33~~8.59 & ~54~~8.28~~8.59 & ~~9~8.34~~8.34 & bl      \\
 &4651.5 &        "                         &$-$0.63$^{B+}$& 29.53        & 113~~8.44~~~9.0&               bl  & ~43~~8.39~~8.54     & ~35~~8.28~~8.54 &   bl             & bl      &  \\
 &4663.5 &$(^{3}P^{0})3s\,^{3}P^{0}~-~ (^{3}P^{0})3p\,^{3}P$&$-$0.61$^B~$&38.22 & ~13~~8.17~~8.44&~18~~8.27~~8.49 & ~~5~~8.37~~8.40   &  ...             &   ...             & ...      &  \\
 &4665.9 &        "                         &$-$0.03$^B~$&38.22         & ~37~~8.16~~8.60 & ~50~~8.24~~8.59 & ~13~~8.38~~8.62     &  ...             &   ...             & ...      &  \\
 &5253.6 &$(^{3}P^{0})3s\,^{3}P^{0}~-~ (^{3}P^{0})3p\,^{3}S$&$-$0.71$^B~$& 38.22  & ~~9~~8.29~~8.39 & ...               &      ...          & ...              &   ...             & ...      & \\
 &5272.5 &"                 &               $-$0.49$^B~$ &38.23&~19~~8.32~~8.42 &~14~~8.28~~8.28 &      ...          & ...              &   ...             & ...      & \\  
  &5695.9 &$3p\,^{1}P ~-~ 3d\,^{1}D$        &~0.02$^B~$&32.10             & ~82~~8.22~~8.29 & bl               & ~10~~8.34~~8.34 & bl                 &   ...             & ...      &  \\
  &6731.0 &$(^{3}P^{0})3s\,^{3}P^{0}~-~ (^{3}P^{0})3p\,^{3}D$&$-$0.29$^B~$&38.22 &~11~~8.37~~8.18 & ...               & ...           &  ...             &   ...             & ...      &  \\
 &6744.3 &        "                         & $-$0.02$^B~$&38.23          & ~11~~8.33~~8.09 & ...               & ...               &  ...             &   ...             & ...      & \\
  &8500.3 &$3s\,^{1}S ~-~ 3p\,^{1}P$       & $-$0.48$^B~$ & 30.64       & ~84~~8.21~~8.59 & 145~~8.40~~8.65       & bl               &  bl             &   ...             & ...      &  \\[1mm]  
  \ion{C}{iv}&5801.3 &$3s^{2}S~-~ 3s^{2}P^{0}$    &$-$0.19$^A~$&37.55   & ~13~~8.34~~8.59 & ~53~~8.45~~8.84 & ...               & ...              &   ...             & ...      & \\  
             &5811.9 &$3s^{2}S~-~ 3s^{2}P^{0}$    &$-$0.49$^A~$&37.55   & ~12~~8.34~~8.59 & ~34~~8.45~~8.84 & ...               & ...              &   ...             & ...      & \\   
 \hline 
\end{tabular}

\noindent bl: blend with other metal line;
{\ldots} : too weak or absent;
XX: LTE does not reproduce the emission;\\
 $\log gf$-values in formal solution: Wiese et al. (1996, 
uncertainties within: 3\% A, 10\% B, 25\% C) and Nahar (2002a, uncertainties within 10-25\% N)
 \end{landscape}

Comparisons between both methods can give discrepancies up to 10-20\% in the 
worst case,
mostly for weak lines in spectral regions where the continuum is difficult to be set, 
the S/N is slightly degraded or there are partial blends with other lines.

One case requires further discussion: the 
\ion{C}{ii} $\lambda\lambda$\,6578/82\,{\AA} doublet in $\tau$\,Sco. 
The remaining problems may be an artifact from the data
reduction. The location of the red wing of H$\alpha$ 
coincides with a bad column of the CCD of FEROS (the spectral orders are
oriented along columns in this spectrograph). A perfect correction for this
cannot be provided. Different wavelengths are affected, depending on the
radial velocity shifts of the object. Note that the region around
\ion{C}{ii} $\lambda\lambda$\,6578/82\,{\AA} doublet has been normalised
relative to the local continuum in Fig.~\ref{fits_tsco}. 

Finally, even when non-LTE-insensitive lines 
are considered in the analysis, it is mandatory to verify that the
ionization balance has been established. This is in order to avoid
systematic errors in the atmospheric parameters 
(in particular $T_{\rm eff}$) and therefore in abundance.
It is recommended that such lines in LTE be used for abundance analyses when an
optimised model atom is not available and the S/N of the observed
spectra is high enough to measure them. Table~3 may act as a
guideline to identify appropriate transitions, depending on atmospheric parameters.

\section{Comparison with previous work}\label{comparison}

We concentrate here on the main sources of systematic uncertainties that can bias
abundance analyses, as identified in Sects.~\ref{atomic_sensitivity} \& \ref{param_variations}: 
atomic data, i.e. different model atoms, and stellar parameters,
i.e. in particular $T_\mathrm{eff}$ scales. 
A more comprehensive comparison is too complex to be performed because
of the large number of variables involved, which in some cases are not even
documented. This includes numerous factors related to observation (e.g.
quality of the analysed spectra, continuum rectification, equivalent width
measurements, line blends), model atmospheres (e.g. codes, abundance standards and linelists considered for line blanketing) and line-formation calculations (e.g. handling of
line blocking, oscillator strengths, line broadening).
A comparison of our results on carbon abundances in the solar neighbourhood
with previous studies will be made in an extra section, because of the wider
implications.

\subsection{Predictions from different non-LTE model atoms}

\begin{figure}[ht!]
\centering
\includegraphics[width=0.78\linewidth]{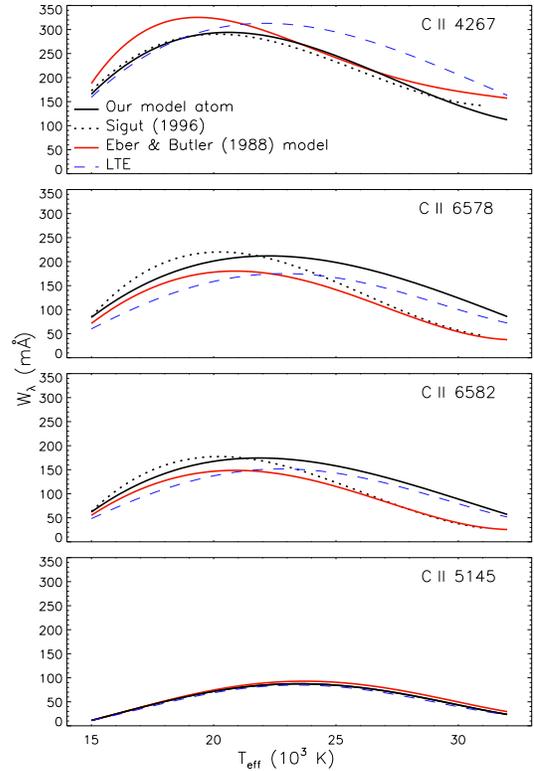}
\caption{Predicted equivalent widths of some \ion{C}{ii} lines from different 
approaches as a function of $T_\mathrm{eff}$. 
All calculations were performed for $\log g$\,$=$\,4.0, $\xi$\,$=$5\,km\,s$^{-1}$
 and $\varepsilon\rm{(C)}$\,$=$\,8.55,
in order to facilitate a comparison with Sigut~(\cite{Sigut96}).
}
\label{EWvsT}
\end{figure}

We have computed non-LTE line profiles based on the present model atom of reference 
and the Eber \& Butler~(\cite{ebbut88}, EB88) model and have compared
them with Sigut's~(\cite{Sigut96}) non-LTE data and our LTE results in Fig.~\ref{EWvsT}.
All models are calculated for the same set of atmospheric parameters and carbon abundances.
Note that only one fundamental difference exists: our computations account for 
non-LTE populations for hydrogen while Sigut assumes LTE. This is one of the reasons why Sigut~(\cite{Sigut96}) found problems 
in reproducing observed trends for the \ion{C}{ii} $\lambda\lambda$\,6578/82\,{\AA} 
doublet at $T_\mathrm{eff}$\,$>$\,25\,000\,K: H$\alpha$ was assumed to be formed
in LTE in that work. 

We obtain consistent abundances from the application of our model atom of
reference to observations for the \ion{C}{ii} $\lambda$\,4267\,{\AA}
transition, as shown in Sect.~\ref{sect_results}. Consequently, the good agreement with
Sigut's predictions indicates that his model atom is also highly useful for 
abundance determinations from this line. Note that Sigut compares his
results (for $\varepsilon$(C)\,$=$\,8.55) only qualitatively with observed 
equivalent widths from different sources, using stellar parameters derived
in one of that studies, or from a photometric $T_\mathrm{eff}$-calibration. 
A reduction of carbon abundance and an improved determination of atmospheric parameters 
as proposed in the present work may bring observation in much better
agreement with his predictions (his Fig.~1). In the region 
$\sim$22\,000\,K\,$\le$\,$T_\mathrm{eff}$\,$\le$\,28\,000\,K good agreement
is also found with the predictions of the EB88 model atom. However, this
model predicts too large equivalent widths and therefore lower abundances
outside this region.

Our analysis gives abundances from the $\lambda\lambda$\,6578/82\,{\AA}
doublet consistent with those from other lines ($\tau$\,Sco may be an exception,
as discussed above). The trends predicted with our model atom of reference 
differ from the three other model calculations. Sigut's and EB88's non-LTE predictions
agree qualitatively up to $T_\mathrm{eff}$\,$\sim$\,22\,000\,K, Sigut's model 
indicating larger and EB88 smaller
equivalent widths, respectively. However, our model shows non-LTE
strengthening throughout, approaching LTE at the high
$T_\mathrm{eff}$-limit, while the two other models imply pronounced
non-LTE weakening at $T_\mathrm{eff}$\,$\gtrsim$\,25\,000\,K. 
Note that this doublet experiences significant non-LTE effects
only when the lines are strong, e.g. for high values of carbon abundance
such as those discussed in Fig.~\ref{EWvsT}. The non-LTE effects are
abundance-dependent, they are reduced with decreasing abundance.
Finally, the predicted equivalent widths for the widely used \ion{C}{ii} multiplet 
including $\lambda$5145\,{\AA} are practically independent of the model assumptions. 
The lines are close to LTE. Any inconsistencies of carbon abundances based
on this multiplet have to be related to other effects, but not by the choice
of the model atom. This multiplet, of the quartet spin system, was not
analysed by Sigut.\\[1cm]

\subsection{Effective temperatures}\label{sect_teff}

Abundance determinations can be systematically biased even in when
a realistic non-LTE model atom is available. The atmospheric parameters need
to be derived accurately as well. We have quantified the effects of
inappropriately chosen atmospheric parameters in Sect.~\ref{param_variations}.
Largely discrepant abundances are found from different ionization stages
in particular for inaccurate effective temperatures.
We were confronted with this from the beginning of our work, when adopting
effective temperatures derived from common photometric calibrations. It was
not possible to establish the ionization equilibrium of \ion{C}{ii/iii} even for 
the lines in LTE
in that case. This motivated us to use an iterative approach to constrain
all variables that are involved in the analysis simultaneously and in a
self-consistent way.

\begin{figure}[t!]
\centering
\includegraphics[width=0.87\linewidth]{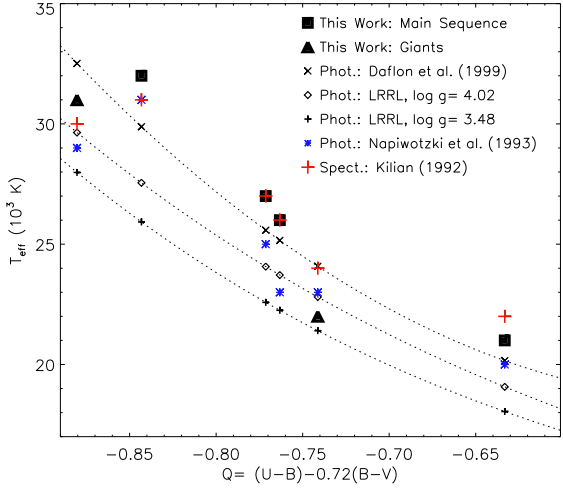}
\caption{Comparison of our final $T_\mathrm{eff}$ for the sample stars
with values derived from photometric calibrations: $T_\mathrm{eff}(Q)$ 
(Daflon et al.~\cite{daf99}; Lyu\-bimkov et al.~\cite{lyu02}: LRRL) 
and $T_\mathrm{eff}([u-b])$ 
(Napiwotzki et al.~\cite{napi93}). Photometric indices are computed
from SIMBAD data. Spectroscopic temperatures according to Kilian~(\cite{k92}) 
are also displayed.
}
\label{teff}
\end{figure}

The final values for $T_\mathrm{eff}$ are compared with other derivations in
Fig.~\ref{teff}. The comparison concentrates on widely used photometric
$T_\mathrm{eff}$-calibrations, which provide a fast determination of
atmospheric parameters, allowing the analysis of larger samples of
stars to be performed. This includes methods using broad-band Johnson photometry, like 
the reddening-free $Q$ index (Daflon et al.~\cite{daf99}; 
Lyubimkov et al.~\cite{lyu02}) or small-band Str\"omgren photometry, as the
$T_\mathrm{eff}([u-b])$ calibration of Napiwotzki et al.~(\cite{napi93}).
The required Johnson and Str\"omgren magnitudes for the individual stars 
are adopted from the SIMBAD database.

The discrepancies in $T_\mathrm{eff}$ can be large, amounting up to $\sim$4000\,K, much
larger than the offsets studied in Sect.~\ref{param_variations}. The
differences in abundance from \ion{C}{ii} and \ion{C}{iii} can then achieve
$\sim$0.7 dex, similar to what is found in previous studies. Our $T_\mathrm{eff}$-values are typically
higher than those from the photometric estimates. The present sample 
is too small to facilitate an improved empirical $T_\mathrm{eff}(Q)$-calibration
on its own. However, there is an indication that giants have 
cooler atmospheres than dwarf stars at the same photometric index.
This is qualitatively in agreement with the findings of Lyubimkov et al.~(\cite{lyu02}).

Spectroscopic $T_\mathrm{eff}$-determinations by Kilian~(\cite{k92}) are
also shown in Fig.~\ref{teff}.
The approach is conceptually the
most similar to the present study despite considerable progress made over the past 15\,years and, as a consequence, gives the closest agreement with our values (except for the cool giant).

These comparisons show that the effective temperatures derived from available 
photometric cali\-brations in early B-type 
stars should be considered only as estimates. 
If adopted as final temperatures, one should be aware of the potential 
systematic effects on the abundance determination (see Sect.~\ref{param_variations}).
Instead, these could be used as
starting points for further refinements via
well-understood spectroscopic indicators.
 This is in order to bring the
model into agreement with all available indicators simultaneously: 
e.g. multiple 
H and He lines and metal ionization equilibria,
as in the present approach. All the atmospheric parameters could then be tightly
constrained. 
Such a step is essential if the 1$\sigma$-uncertainties in the abundance 
analysis are required to be smaller than 
$\sim$0.3 dex.

\section{The stellar present-day C abundance in the solar
neighbourhood}\label{neighbourhood}

\begin{figure}
\centering
\includegraphics[width=\linewidth]{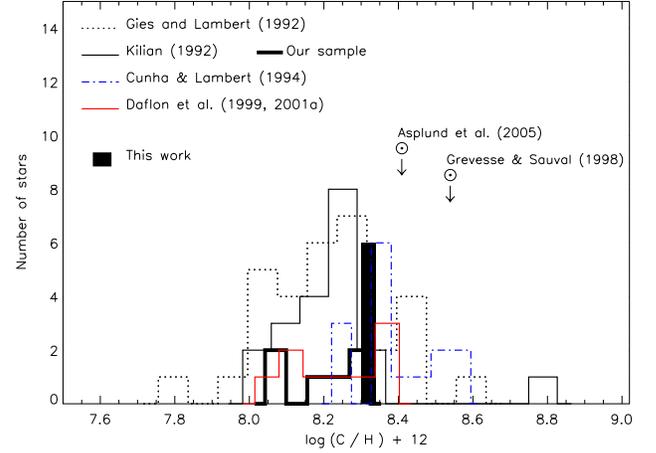}
\caption{Comparison of C abundances derived from our sample of 
early B-type stars in the solar vicinity (from the field and from OB
associations) with results from the literature. Similar objects 
at distances shorter than $\sim$1\,kpc from the Sun and at galactocentric
distances within up to $\sim$500\,pc difference with respect to the location of
the Sun are considered. The most recent solar values are also indicated.
The binsize of the histograms is related to the statistical uncertainty
of each sample. Our sample is not large enough for a statistical comparison,
therefore a column with a thickness of our uncertainty is displayed.
The programme stars coincide with six objects from Kilian~(\cite{k92}): her 
abundances show a much larger spread for these. See the text for details
}
\label{histogram}
\end{figure}

Early-type stars can act as tracers for the present-day chemical composition in the
solar neighbourhood.
The present small sample of stars is suited to address this
topic, as the objects are randomly distributed over nearby OB associations (three
stars in Ori\,OB1 and Sco-Cen, see Table~\ref{parameters}) and the field (the other stars).
They are located at distances 
shorter than $\sim$1\,kpc from the Sun and at galactocentric distances 
within up to $\sim$500\,pc difference with respect to the location of the Sun.
We find a highly-homogeneous carbon abundance of 
$\varepsilon$(C)\,$=$\,8.32$\pm$0.04 (1$\sigma$ statistical uncertainty). 
Continuing the discussion from the last section,
the carbon abundance derived in this work is compared with 
results from selected previous studies of early B-type dwarf and
giant stars in non-LTE, as shown in Fig.~\ref{histogram}.

Non-LTE analyses based on the EB88 model atom, 
using different linelists, find systematically higher abundances than previous
LTE studies. However, a large spread in abundance is still present in each sample. 
Our results can be directly compared with those of the purely spectroscopic
study of Kilian~(\cite{k92}). 
A mean $\varepsilon$(C)\,$=$\,8.19$\pm$0.12 is derived
from the same six stars when adopting her results. 
This comparison is highly important as it shows that a self-consistent analysis, 
like the one presented here, may
drastically reduce the statistical scatter and may also imply a considerable
systematic shift of the mean abundance.
Similar to Kilian, Gummersbach et al.~(\cite{Gummersbachetal98}, not
shown in Fig.~\ref{histogram}) also performed 
a consistent spectroscopic analysis of five stars in the solar 
vicinity as part of their Galactic abundance gradient study.
Their results also indicate systematically lower 
abundances and a considerable spread.

Examples of non-LTE analyses for larger samples of stars based on effective 
temperatures derived from photometric calibrations are also displayed in 
Fig.~\ref{histogram}. 
Gies \& Lambert~(\cite{gl92}) and Cunha \& Lambert~(\cite{cl94}) derived
$T_\mathrm{eff}$ from calibrations based on Str\"omgren photometry and non-LTE 
abundances from the \ion{C}{ii} multiplet around
$\lambda$5145\,{\AA} and where possible from \ion{C}{ii}   
$\lambda\lambda$5648/62\,{\AA}. Both multiplets originate in the
quartet spin system and they are almost unaffected by non-LTE effects.
For the comparison we have excluded the five supergiant stars of the sample
of Gies \& Lambert~(\cite{gl92}). 
Gies \& Lambert~(\cite{gl92}) have two stars in common with the present sample, indicating
systematically lower abundances: for HR\,2928 $\Delta\varepsilon(\mathrm
C)$\,$\simeq$\,$-$0.3\,dex and for HR\,1861
$\Delta\varepsilon(\mathrm C)$\,$\simeq$\,$-$0.2\,dex 
(from the same lines). They also consider the 
doublet \ion{C}{ii} $\lambda\lambda$6578/82\,{\AA} for several stars of their
sample, but obtain even lower abundances compared to the lines from
the quartet spin system (by approximately another $-$0.3\,dex).
Cunha \& Lambert~(\cite{cl94}) have only one star in common with us, HR\,1861. 
They found a similar abundance for this star as Gies \& Lambert~(\cite{gl92}). 
More recently, Daflon et al.~(\cite{daf99}, \cite{daf01}) derived temperatures
from their calibration of the Johnson $Q$-parameter and non-LTE 
abundances from the \ion{C}{ii} multiplet around $\lambda$5145\,{\AA}. 

All the data on carbon abundances from early-type stars summarised above was 
{\em derived from interpretation of observation}, which may be affected by many
sources of systematic error (see Sect.~\ref{calib}). These data are {\em interpreted in turn to
 test models} of massive star evolution and
 the chemical evolution of the Galaxy, which are anchor
points for studies of stellar and galactochemical evolution
in general. 
Any misinterpretation in the first step in this chain may have severe
consequences for our understanding of the whole picture. 
In this context, it is important to investigate
the conclusions drawn from a diagram like Fig.~\ref{histogram}.

\addtocounter{table}{1}
\begin{table}
\caption{Carbon abundances of different objects in the solar vicinity} 
\label{Cneighbourhood}
\begin{tabular}{lcl}
\hline
Objects                        & $\varepsilon$(C) & Source\\
\hline
B stars (pristine value)       & 8.35$\pm$0.05 & present work\\[.5mm]
B stars                        & 8.25$\pm$0.08 & Herrero~(\cite{Herrero03})\\
Orion \ion{H}{ii} (gas)        & 8.42$\pm$0.02 & Esteban et al.~(\cite{Estebanetal04})\\
Orion \ion{H}{ii} (gas$+$dust) & 8.52$\pm$0.02 & Esteban et al.~(\cite{Estebanetal04})\\
young F and G stars            & 8.55$\pm$0.10 & Sofia \& Meyer~(\cite{SoMe01})\\
ISM                            & 8.15$\pm$0.06 & Sofia \& Meyer~(\cite{SoMe01})\\[.5mm]
Sun                            & 8.39$\pm$0.05 & Asplund et al.~(\cite{ags05})\\
Sun                            & 8.52$\pm$0.06 & Grevesse \& Sauval~(\cite{gs98})\\
\hline
\end{tabular}
\end{table}

 Previous non-LTE studies 
all show a broad range of carbon abundances in the solar
neighbourhood, spanning a factor of $\sim$10 in total.
Current models of stellar evolution and chemical evolution of the Galaxy do not provide a physical explanation to the large range shown in Fig.~\ref{histogram} for this kind of star located in the solar vicinity\footnote{When considering also LTE results
(e.g. Kane et al.~\cite{kane80}) the whole abundance range rises up to 1.6 dex (a factor of 40).}.
If this extended distribution were real, its physical interpretation may require 
 to account for the following physical processes.
{\sc i})  Mechanisms that alter the atmospheric structure, such as
sufficiently strong magnetic fields or diffusion/radiative levitation processes 
acting on short timescales and independent of atmospheric parameters and evolutionary age.
These are so far discussed only for a few cases, well motivated by
observational evidence. A common occurrence of these processes would imply that the
basic assumption of homogeneity of the atmosphere may be invalid and as a consequence
the modelling techniques applied so far may be inadequate.
{\sc ii}) Extremely efficient depletion mechanisms in the course of
stellar evolution, by one to two orders of magnitude larger than currently
predicted (e.g. Meynet \& Maeder~\cite{mm03}). Already on the main sequence
mixing with matter from the stellar core would require a higher efficiency 
than currently predicted for {\em convection} during the first dredge-up.
Eventually, also enrichment mechanisms may be required which contradict
nucleosynthesis (the 3$\alpha$-process is not active in OB dwarfs and
giants, carbon is instead depleted by the CNO cycle).
{\sc iii}) An enormous chemical inhomogeneity of the present-day interstellar 
material in the solar neighbourhood out of which
the stars have been formed. This would require the possibility to change abundances 
practically instantly even within single clusters by amounts that are otherwise 
attributed to the past $\sim$12\,Gyrs of Galactochemical evolution 
(e.g. Chiappini et al.~\cite{Chiappinietal03}).

Concerning the chemical inhomogeneity, non-uniform
abundance distributions as a result of galactochemical evolution are also 
discussed in literature. E.g., Oey~(\cite{Oey03}) indicates a scatter of
0.13\,dex in oxygen abundance for a population at solar metallicity, which
is in reasonable agreement with previous observational constraints. 
However, this is an upper limit, as homogenisation 
processes were neglected entirely, such that the true scatter can be expected to be
much lower.

The present work, in combination with Paper~1 and 2, avoids these
fundamental problems. It implies homogeneous abundances after 
improving the modelling {\em and} the analysis methodology, bringing all model aspects into 
agreement with observation at once, which was not achieved so far.

{\sc i}) It conforms with the finding of a uniform gas-phase carbon abundance in the
ISM (out to distances of 1.5\,kpc, e.g. Sofia \& Meyer~\cite{SoMe01}, and
references therein), with the systematic offset between the two values
being a consequence of dust formation in the ISM.
The uncertainties of the ISM abundance (see Table~\ref{Cneighbourhood}) are
determined by the accuracy to which the oscillator strengths of the
resonance lines in the UV are known, and should not exceed more than
$\sim$10\%. 

{\sc ii}) It also agrees with Galactochemical evolution
models, which predict homogeneous abundances in the solar neighbourhood
(e.g. Chiappini et al.~\cite{Chiappinietal03}). A variety of hydrodynamic
processes should keep the ISM chemically well-mixed on small time-scales
(Edmunds~\cite{edmunds75}; Roy \& Kunth~\cite{rk95}). The variation due to the
Galactic abundance gradient should amount to up to
$\sim$0.04--0.08\,dex\,kpc$^{-1}$,  which is of the order of the uncertainty found here (one kpc is the maximum Galactocentric distance sampled in our comparison). 

{\sc iii}) Another aspect concerns the notably sub-solar abundances  
from early-type stars found so far (e.g. Herrero~\cite{Herrero03}, with respect to
the old solar standard, Grevesse \& Sauval~\cite{gs98} -- see
Table~\ref{Cneighbourhood}). 
The present findings remedy the situation, in particular
if accounting for the recently revised solar carbon abundance (Asplund et
al.~\cite{ags05}, see also Table~\ref{Cneighbourhood}). 

{\sc iv}) Finally, rotationally-induced mixing with CN-processed material from the core may 
change the atmospheric composition. 
Theory predicts
a depletion of carbon by $\sim$0.03\,dex for a star of 20 M$_\odot$ with 
initial rotational velocity 300\,km\,s$^{-1}$ evolving from the zero-age
Main Sequence to the end of the Main Sequence stage, reaching
$\sim$0.15\,dex in the supergiant stage (e.g. Meynet \& Maeder~\cite{mm03}).
The presence of a magnetic field may amplify rotational mixing
(Maeder \& Meynet~\cite{mm05}), however the effects on abundances are not
expected to exceed a factor $\sim$2. As absolute rotational velocities of
stars can be measured only for a few exceptionally cases, this topic can be
addressed comprehensively only by a statistical approach. Our sample of stars 
is not large enough for this. We note however, that we do not find a significant 
trend of carbon abundances
with evolutionary age for these apparently slow-rotating stars (only
$\tau$\,Sco is suggested to be a {\em real} slow rotator, Donati et
al.~\cite{donati06}). The abundances derived may not correspond to the pristine values 
nonetheless, but it is unlikely that all the stars are very fast rotators
seen pole-on.
Small corrections by up to $\sim$$+$0.05\,dex per star (depending on
evolution stage) are predicted by theory for objects of average rotation.
We estimate a 
correction of $+$0.03\,dex to the average value
of our sample (considering a small increase of the uncertainty),
in order to derive an average pristine abundance of $\varepsilon(C)=8.35 \pm 0.05$.\\[-2mm]

The average pristine carbon abundance from early B-type stars in the solar
neighbourhood is compared with other indicators in Table~\ref{Cneighbourhood}.
The value is not only in agreement with the revised solar abundance but also
with the gas-phase abundance derived for the Orion nebula. The finding of
a uniform abundance matches with a homogeneous ISM abundance. A disagreement
is found only with abundances from young F \& G-type stars. This may be
resolved when hydrodynamic 3D-analyses of late-type stars become routine, as
the discrepancy is similar to the difference between the old and the revised
solar standard, which is the result of such an improved analysis. The
conclusion of Sofia \& Meyer~(\cite{SoMe01}) that B-type stars are not reliable
proxies for present-day abundances in the solar neighbourhood may need
a revision in view of the present results.  

Despite the small sample size analysed so far, our accurate results
indicate that the sub-solar average value and the large scatter 
of C abundances in early-type stars found in previous
non-LTE studies could be mostly a consequence of systematic uncertainties.
These may be easily introduced by the choice of inappropriate atomic data 
and/or stellar parameters, as we have shown. However, a confirmation of our findings 
from analysis of a larger sample of stars is required, also for other elements.

\section{Summary and Conclusions}

The motivation of this work was the solution of a long-standing problem 
in stellar astrophysics: 
the reliable determination of carbon abundances from early-type stars. 
For this purpose we constructed a sophisticated \ion{C}{ii/iii/iv} model atom for
non-LTE line-formation calculations based on input atomic data that were carefully
selected by an empirical calibration process. This was performed through an
extensive iteration scheme that not only allowed us to constrain the
input atomic data but also simultaneously the atmospheric parameters of our
calibration stars, the basis for all further studies. 
The calibration sample consists of six bright and apparently
slow-rotating early B-type stars in the solar neighbourhood, with high-S/N, high-resolution spectra and a broad wavelength coverage. 

The self-consistent analysis provides atmospheric parameters with
unprecedented accuracy and with reduced systematic error: for
effective temperature the uncertainties are as low as $\sim$1\% and for surface gravity $\sim$10\%.
The atmospheric parameters derived from the carbon ionization equilibria
are highly consistent with our previous spectroscopic analyses:
results from a non-LTE study of H and He in the visual and in the
near-IR, including the establishment of the \ion{He}{i/ii}
ionization equilibrium in the hotter stars (Paper~1), and the reproduction
of the spectral energy distributions from the UV to the near-IR (Paper~2).

All C lines considered, from up to three ionization stages,
indicate similar abundances. 
The linelist includes 40 transitions suitable for analysis over a wide wavelength
range. In particular, the strongest features, of highest importance for extragalactic applications, are consistently modelled. 
The statistical 1$\sigma$-uncertainties  
of the C abundance in each star are of the order $\sim$0.05-0.10\,dex. We
estimate the systematic 1$\sigma$-uncertainties to be $\sim$0.10-0.15\,dex.
The average carbon abundance from \ion{C}{ii/iii/iv} in the sample stars is highly
uniform, $\varepsilon$(C)\,$=$\,8.32$\pm$0.04. 

These results of unprecedented precision 
provide important constraints for stellar evolution and the
chemical evolution of the Galaxy, despite the small sample size:
{\sc i}) They suggest that carbon depletion due to rotational mixing in the
course of stellar evolution is small for stars without excessive initial
angular momentum (by $<$0.05\,dex) up to the giant stage, in agreement with 
theoretical predictions.
{\sc ii}) In consequence they suggest that the present-day carbon abundance
in the solar neighbourhood is higher and much more homogeneous than indicated by
previous work on early-type stars. This is consistent with the uniform
abundance found in studies of the ISM and with predictions from models of the 
chemical evolution of the Galaxy. Moreover, our re-evaluated stellar value
is in agreement with the gas-phase abundance derived for the Orion
\ion{H}{ii} region and the recently revised solar abundance.  

This work shows that one should not underestimate the
importance of a careful choice of input atomic data for non-LTE analyses. 
Not only accurate data for radiative transitions are required but also for 
collisional transitions. The carbon abundance analysis also turned out to
react highly sensitively to the choice of atmospheric parameters, in
particular to effective temperatures. If not accounted for properly, both 
factors may result in systematic errors in the interpretation of observed
spectra, which cannot be reduced by statistics, i.e. an increase of the
sample size of stars to be analysed. 
Detailed comparisons and calibrations of models with 
observed spectra are of highest importance, as these constitute the 
only empirical constraints for quantitative analyses. 
Only then can stellar and Galactochemical
evolution models be verified in a meaningful way, and the studies be
extended to stars in other galaxies.

\begin{acknowledgements}
The authors wish to thank U.~Heber and K.~Cunha for their interest
and their support of the project. We further thank K.~Butler for making available
{\sc Detail} and {\sc Surface} and M.~Altmann for kindly providing the FEROS data.
M.F.N. thanks S.~Daflon for encouraging the study of this subject 
and also acknowledges a Ph.D. scholarship by the
Deutscher Akademischer Austausch Dienst (DAAD).
\end{acknowledgements}

\Online

\begin{figure*}[!ht]
\centering
\includegraphics[width=0.98\linewidth]{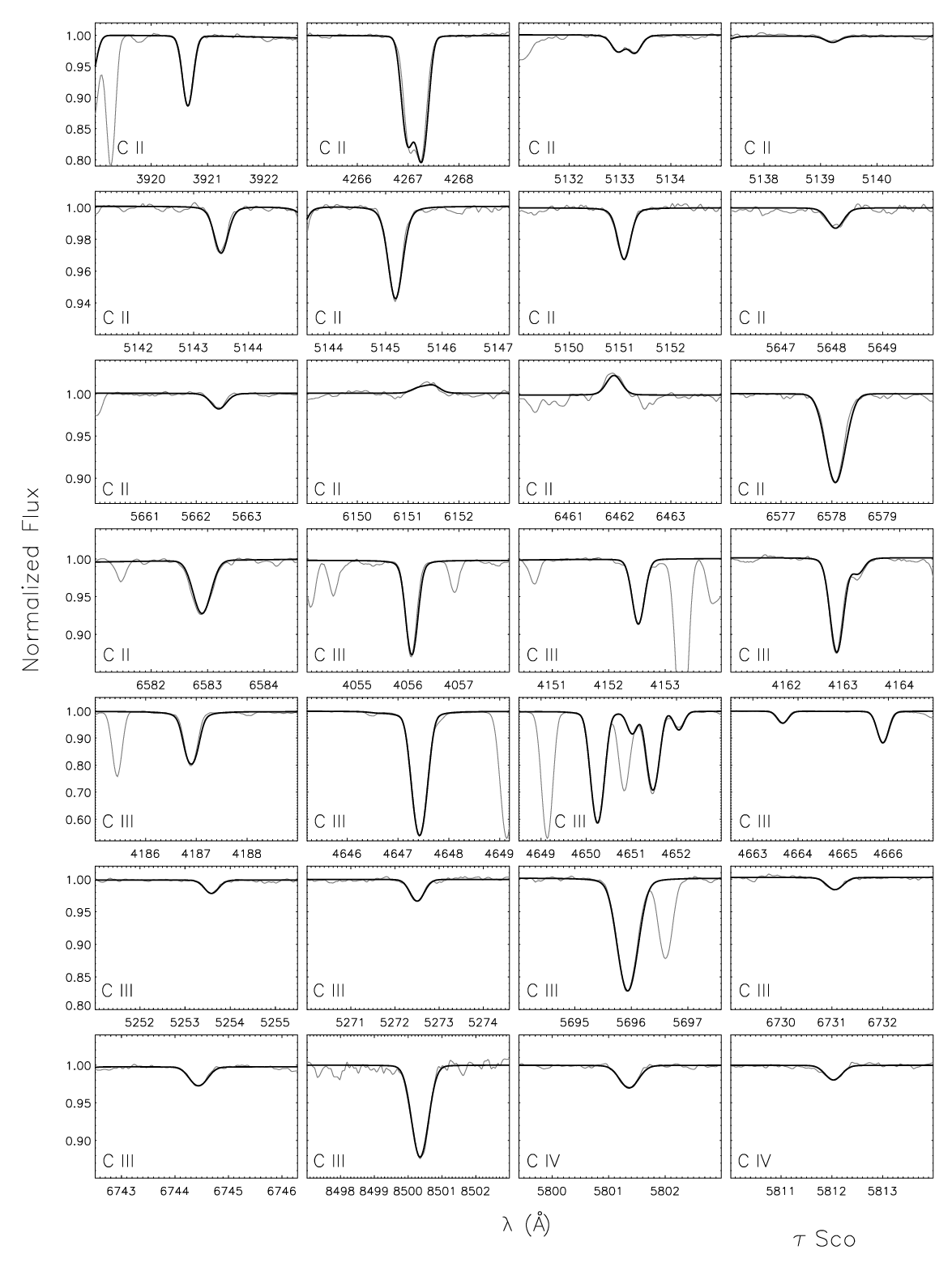}\\[-4mm]
\caption{Examples of our line profile fits (black line) to the 
observed high-quality spectrum (grey line) of $\tau$~Sco. 
}
\label{fits_tsco}
\end{figure*}

\begin{figure*}[ht!]
\centering
\includegraphics[width=0.985\linewidth]{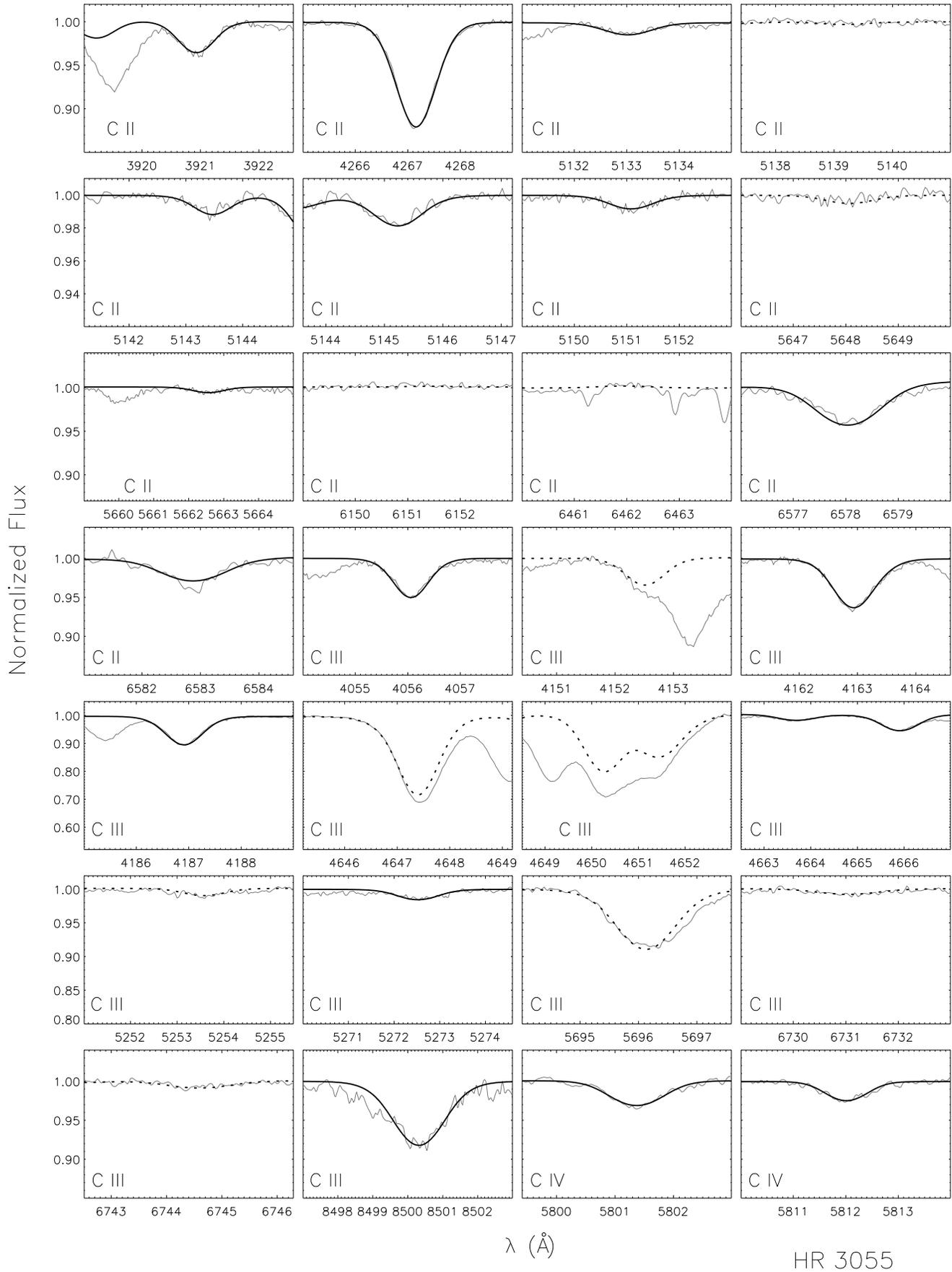}
\caption{As Fig.~\ref{fits_tsco}, but for HR~3055. Dotted lines 
indicate the features not included in the abundance determination.
}
\label{fits_3055}
\end{figure*}

\begin{figure*}[ht!]
\centering
\includegraphics[width=0.985\linewidth]{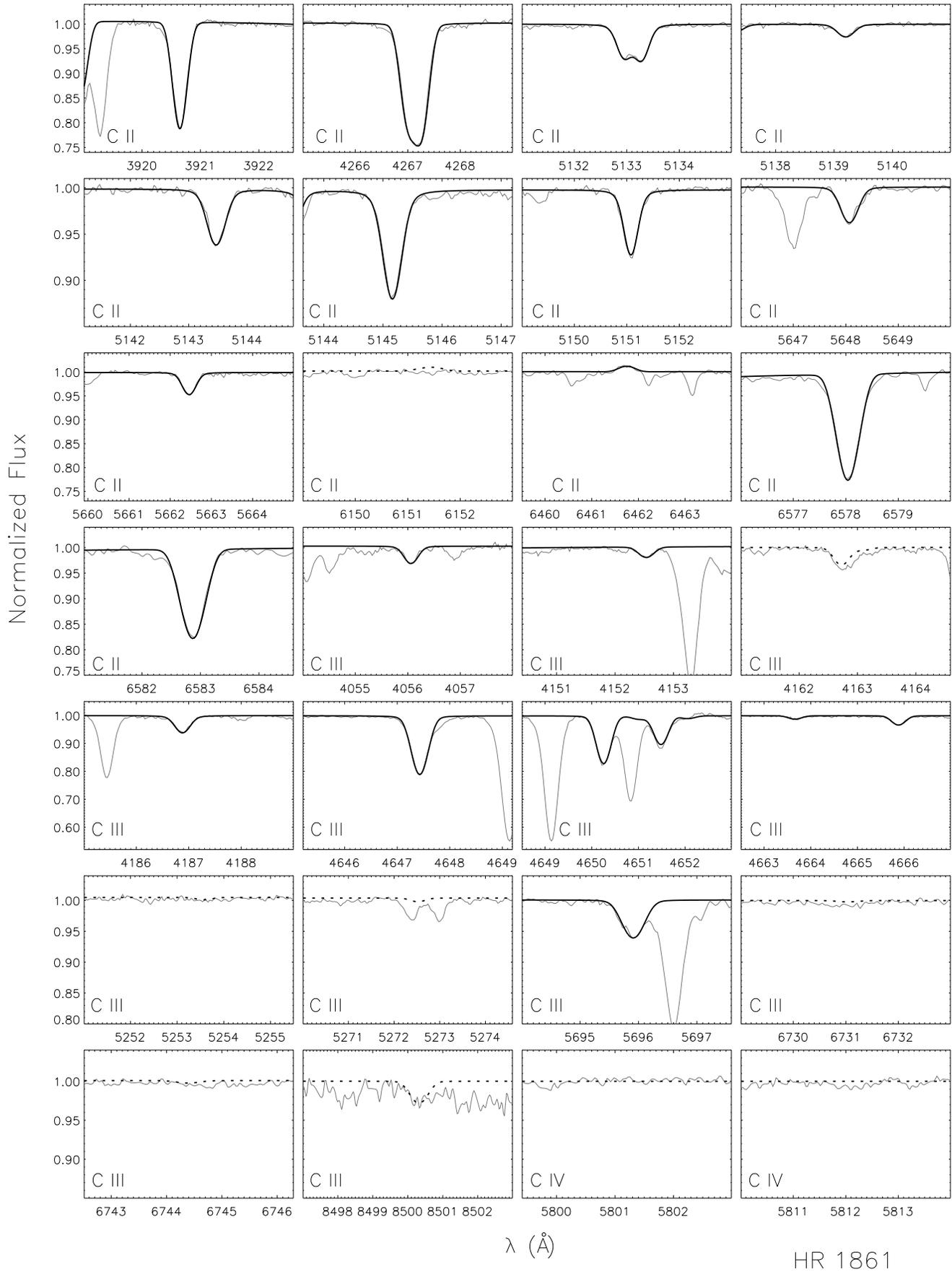}
\caption{As Fig.~\ref{fits_3055}, but for HR~1861. 
}
\label{fits_1861}
\end{figure*}

\begin{figure*}[ht!]
\centering
\includegraphics[width=0.985\linewidth]{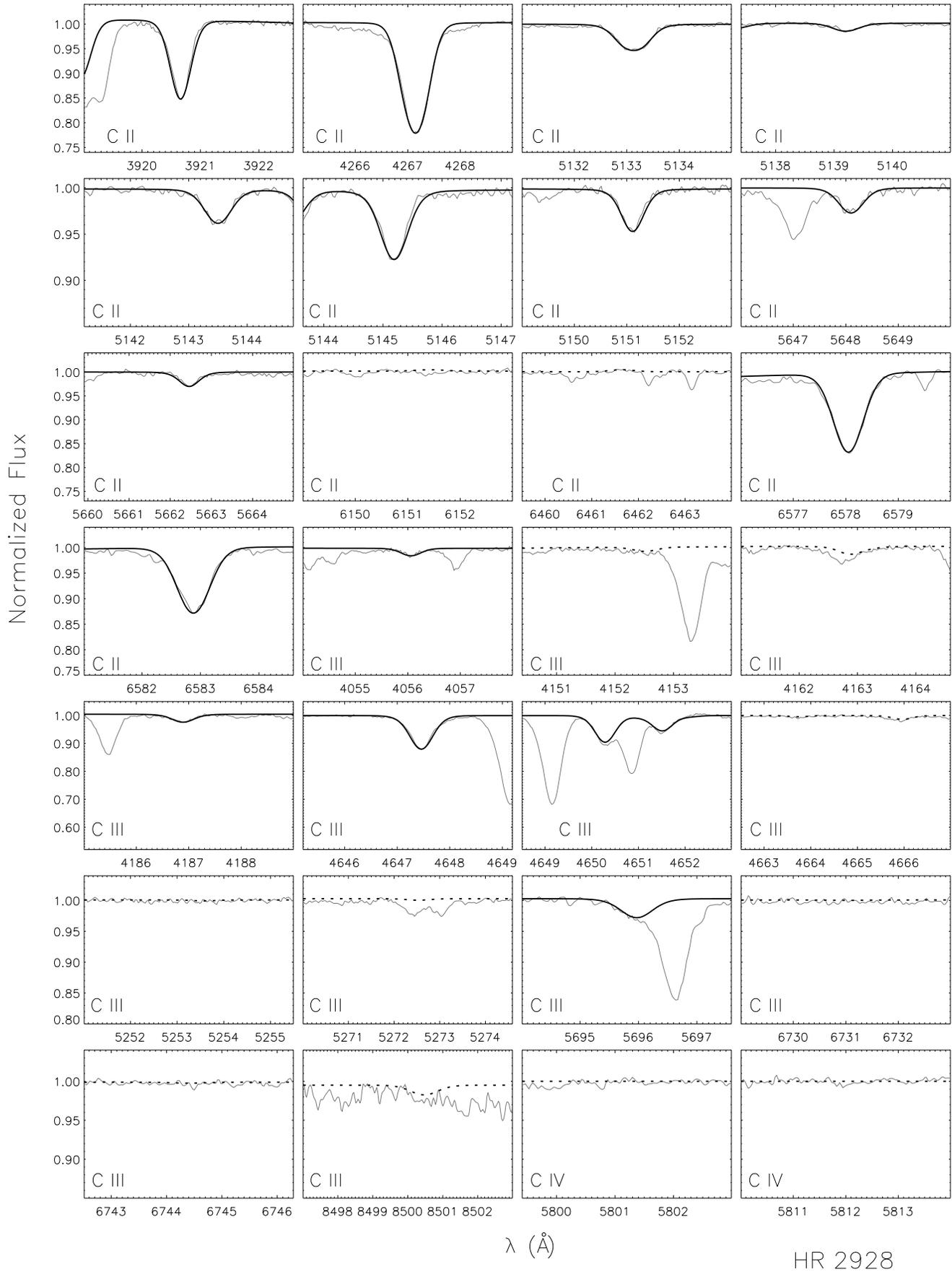}
\caption{As Fig.~\ref{fits_3055}, but for HR~2928. 
}
\label{fits_2928}
\end{figure*}

\begin{figure*}[ht!]
\centering
\includegraphics[width=0.985\linewidth]{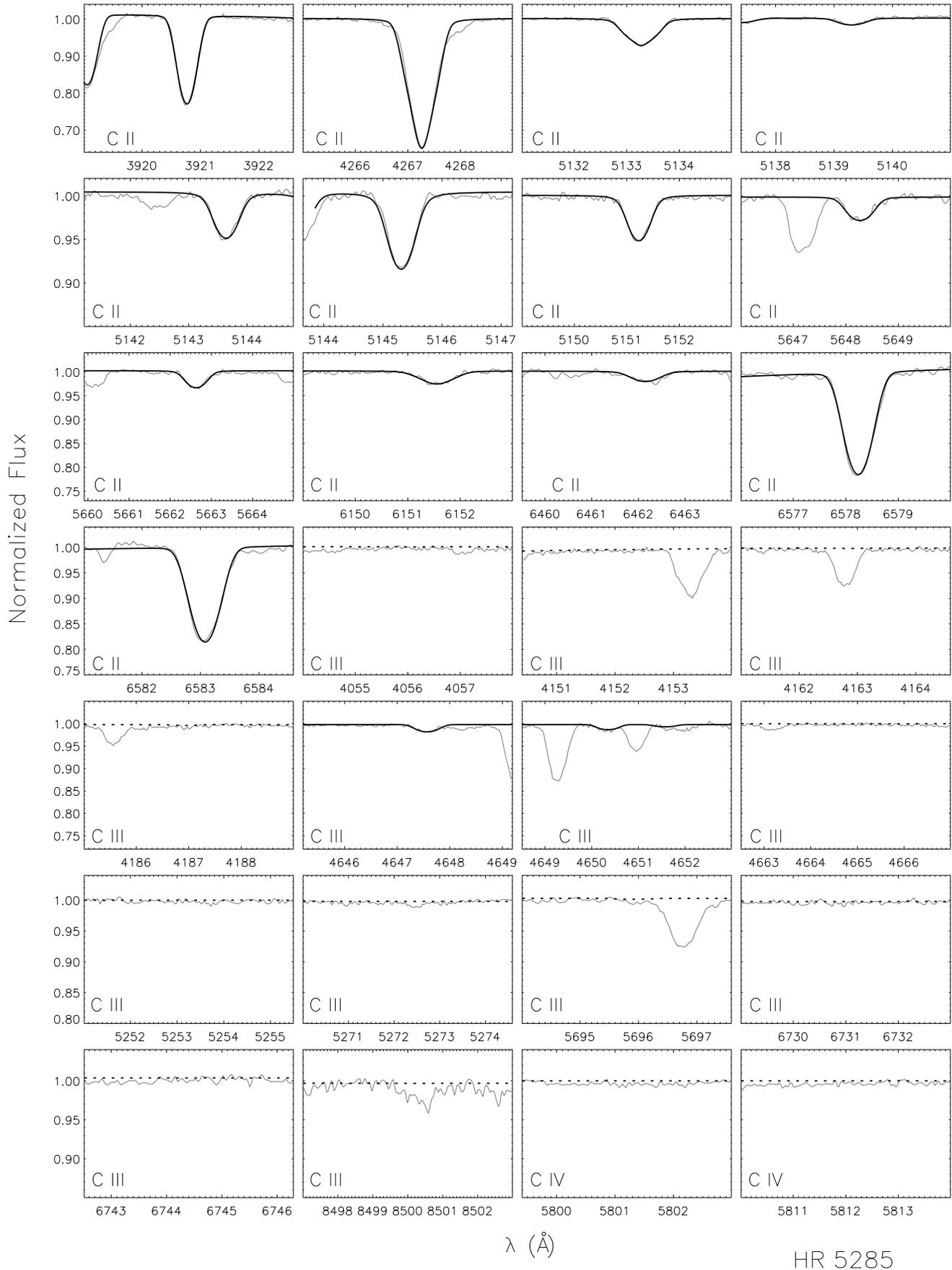}
\caption{As Fig.~\ref{fits_3055}, but for HR~5285. 
}
\label{fits_5285}
\end{figure*}

\begin{figure*}[ht!]
\centering
\includegraphics[width=0.985\linewidth]{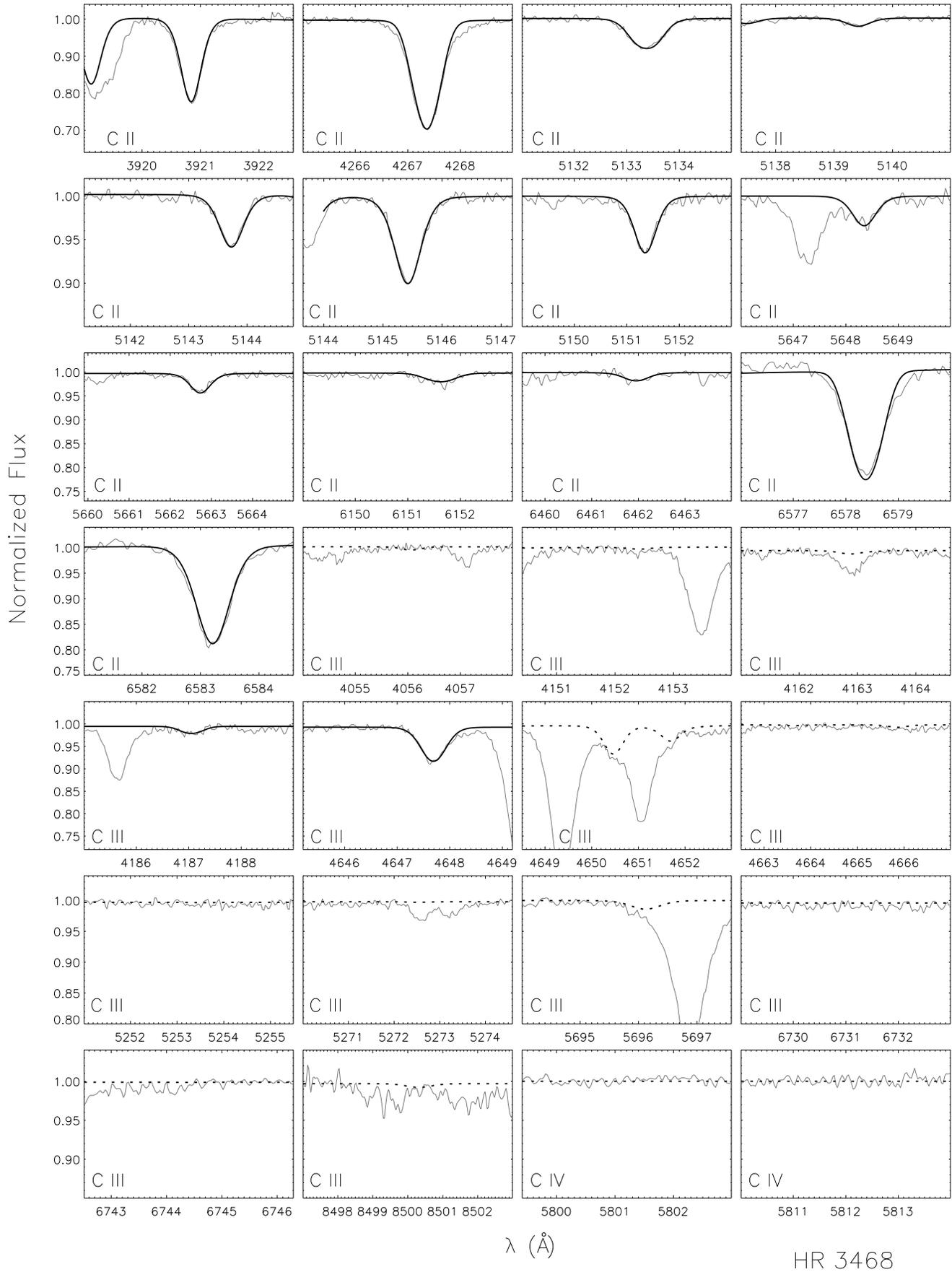}
\caption{As Fig.~\ref{fits_3055}, but for HR~3468. 
}
\label{fits_3468}
\end{figure*}

\end{document}